\documentclass[11pt,a4paper,twoside]{article}
\usepackage{amsmath,amsfonts,amssymb,mathtools}
\usepackage{subdepth}
\usepackage[T1]{fontenc}
\usepackage{ae}
\usepackage{aecompl}
\usepackage[utf8]{inputenc}
\usepackage{cite}
\usepackage{enumitem}
\usepackage{array}
\newcolumntype{L}[1]{>{\raggedright\let\newline\\\arraybackslash\hspace{0pt}}m{#1}}
\newcolumntype{C}[1]{>{\centering\let\newline\\\arraybackslash\hspace{0pt}}m{#1}}
\newcolumntype{R}[1]{>{\raggedleft\let\newline\\\arraybackslash\hspace{0pt}}m{#1}}
\usepackage{fancyhdr}
\usepackage{setspace}
\usepackage[hyperindex,breaklinks]{hyperref}
\usepackage{graphicx}
\usepackage[small,bf,font=small,textfont=it,labelsep=period]{caption}
\usepackage[left=2.8cm,right=2.8cm,top=2.5cm,bottom=3cm]{geometry}
\graphicspath{{figures/}}

\numberwithin{equation}{section}
\def\PC{Poincar\'e}
\def\ISO{$\mathrm{ISO}(2)$}
\def\SO{$\mathrm{SO}(3)$}
\def\SU{\mathrm{SU}(2)}
\def\SL{\mathrm{SL}(2,\mathbb{C})}
\def\CC{\mathbb{C}}
\def\RR{\mathbb{R}}
\def\AA{\mathbb{A}}
\def\BB{\mathbb{B}}
\def\II{\mathbb{I}}
\def\KK{\mathbb{K}}
\def\pp{\mathfrak{p}}
\newcommand{\nn}{\nonumber\\}
\newcommand\st[2]{|#1;#2\rangle}
\newcommand{\ket}[2]{\langle #1,#2\rangle}
\newcommand{\bra}[2]{[ #1,#2 ]}

\newcommand{\Pgen}[3]{\langle#1|\,#2\,|#3]}
\DeclareRobustCommand{\bfrac}[2]{%
	\mathchoice{\frac{\raisebox{-0.4ex}{$#1$}}{\raisebox{-0.3ex}{$#2$}}}%
	{\frac{\raisebox{-0.3ex}{$\scriptstyle#1$}}{\raisebox{-0.4ex}{$\scriptstyle#2$}}}%
	{\frac{#1}{#2}}%
	{\frac{#1}{#2}}%
}
\usepackage[oldsyntax]{stackengine}
\newcommand\hatj{%
	\stackengine{\Sstackgap}{$j$}{\(\hspace{3.8pt}\hat{}\)}{O}{l}{F}{T}{S}}
\begin{document}	%%%%%%%%%%%%%%%%%%%%%%%%%%%%%%%%%%%%%%%%%%%%%%%%%%%%%%%%
%%%%%%%%%%%%%%%%%%%%%%%%%%%%%%%%%%%%%%%%%%%%%%%%%%%%%%%%%%%%%%%%%%%%%%%%
\begin{titlepage}
\thispagestyle{empty}
\topmargin=3cm
\oddsidemargin=17pt
\begin{center}
	\begin{minipage}[t]{1\textwidth}
	\centering\setstretch{2}
	\textsc{\LARGE
		The four-dimensional on-shell three-point}\\
	\textsc{\LARGE
		Amplitude in spinor-helicity formalism}\\
	\textsc{\LARGE
		and BCFW recursion relations}
	\end{minipage}
		
		\vspace{6em}
		
		{\Large Andrea Marzolla$^{\S}$}
		
		\vspace{1.5em}
		\parbox[t]{0.7\textwidth}{\itshape \footnotesize \centering
			{$^{\S}$Physique Th\'eorique et Math\'ematique and International Solvay Institutes, Universit\'e Libre de Bruxelles, C.P. 231, 1050 Brussels, Belgium.}
			}

	\vspace{5.5cm}

{\bf \small \textsc{Abstract}}
\vskip 1ex
	\begin{minipage}[t]{\textwidth}
	\noindent
	Lecture notes on Poincar\'e-invariant scattering amplitudes and tree-level recursion relations in spinor-helicity formalism. We illustrate the non-perturbative constraints imposed over on-shell amplitudes by the Lorentz Little Group, and review how they completely fix the three-point amplitude involving either massless or massive particles. Then we present an introduction to tree-level BCFW recursion relations, and some applications for massless scattering, where the derived three-point amplitudes are employed.
	\end{minipage}
\end{center}

\end{titlepage}

\clearpage{\pagestyle{empty}\cleardoublepage}
\pagenumbering{roman}

%%%%%%%%%%%%%%%%%%%%%%%%%%%%%%%%%%%%%%%%%%%%%%%%%%%%%%%%%%%%%%%%%%%%%%%%
\subsection*{Note to the reader}

The present notes collect and integrate the subjects taught over five hours at the \href{http://www.ulb.ac.be/sciences/ptm/pmif/Rencontres/ModaveXII/lectures.html}{XII Modave Summer School in Mathematical Physics}. The audience was composed almost entirely by young PhD students, so the lectures were intended for researchers who are new to this specific field. In the same way, these notes require as preliminary notions no more than the standard material of any master studies in theoretical physics, namely some basics of group theory, quantum field theory and complex analysis. 

Moreover, the shortness of the course imposed a restricted selection of topics. The study of theoretical and mathematical aspects of on-shell scattering amplitudes is an extremely broad field of research, which experienced great vitality and development in the recent years. This work does not aim to any extent at being an exhaustive review on the subject, reviews which on the other hand we have in a number of eminent examples (as for instance:~\cite{Elvang:2013cua,Dixon:2013uaa,Benincasa:2013faa}).

This text is rather meant to be an initiating tool for students and researchers who are interested in working on these topics. Therefore, taking the risk of being even redundant, we provide enough details for the computations to be reproduced by the reader. On the other hand, we try on any occasion to refer to other material which extends and develops the subjects that we are neglecting in our discussion, in order to guide the interested reader.

\clearpage{\pagestyle{empty}\cleardoublepage}
\pagenumbering{arabic}
%%%%%%%%%%%%%%%%%%%%%%%%%%%%%%%%%%%%%%%%%%%%%%%%%%%%%%%%%%%%%%%%%%%%%%%%
\section*{Introduction\markboth{Introduction}{}}
    \setcounter{section}{0}
    \addcontentsline{toc}{section}{Introduction}

The most prolific source of experimental evidence for high energy physics is given by large particle accelerators, where the measured observables are the probability amplitudes of scattering processes. Such measured quantities are matched with theoretical predictions which are computed mainly by techniques based on Feynman diagrammatics. These techniques are derived from Lagrangian formulation of quantum field theory, and rely over perturbative expansion and renormalizability. They apply thus only to \emph{weak} and \emph{renormalizable} interactions. 

However, since the very beginning of quantum field theory, an alternative non-perturba\-tive analytic approach to scattering amplitudes was suggested~\cite{PhysRev.52.1107,1946ZNatA608H}. Such approach, initially developed by Chew~\cite{RevModPhys.34.394}, and by many others afterwards, was prominent in the 50's and 60's, going under the name of ``$S$-matrix program''. The idea was that the $S$-matrix could be entirely reconstructed starting from a few first principles: \emph{analyticity}, \emph{unitarity} and \emph{crossing} properties of the $S$-matrix, together with its \emph{symmetries}. 

The program, after some initial success, did not manage to achieve its goal, and was eventually supplanted by the field theoretical approach, namely after the affirmation of quantum chromodynamics as theory of strong interaction. It knew then a period of rediscovery during the 70's and 80's, in the framework of two-dimensional perturbative string-theory, but it is in very recent times that the $S$-matrix perspective is living a new revival, more as a complement, rather than a substitute, of perturbative quantum field theory. Ironically, this new \emph{vague} of success is concerning mainly Yang-Mills theories, which were responsible for burying the $S$-matrix program half a century ago. The new elements that made the recent advancements achievable are the focus on massless particles, with the extensive use of the spinor-helicity formalism as a simplifying operational framework, and the addition of more symmetry (supersymmetry, conformal symmetry).

The symmetries of spacetime have a crucial role in the scattering theory, since they underlie the notion itself of particle-state: what we call \emph{elementary particle} in high-energy physics corresponds to a \emph{unitary representation} of the spacetime symmetry group. But spacetime symmetries do not only classify the external states participating in the scattering, they also yield constraints on the form of the scattering amplitude. Such symmetry-based constraints do not rely on the existence of a local Lagrangian, nor on the validity of a perturbative expansion, so they must apply to any kind of particles (any mass, any spin) and interactions (strong, non-renormalizable,...).

In these notes we will remain in the `maximally realistic' situation of scattering events in four-dimensional flat spacetime, and we will show how mere Poincar\'e invariance is enough to fix the kinematic dependency of the \emph{three-point} amplitude, where the external states are either \emph{massless} or \emph{massive}. 

The most general non-perturbative \PC-invariant three-point amplitude is already quite an interesting result on its own, but it is even more significant since it can be used as building-block to construct higher-point amplitudes, through \emph{on-shell recursion relations}. We present here an introduction to tree-level BCFW (Britto-Cachazo-Feng-Witten) recursion relations, and we carry out some practical applications to simple but emblematic examples, with an attentive regard towards the details of computations. 

Furthermore, the entirety of our derivation will be realized in spinor-helicity formalism. The advantages of such formalism, besides being particularly convenient to express on-shell massless amplitudes, are two main ones: in the first place, the constraints coming from Lorentz symmetry take a particularly simple and effective form in spinor language; then, the spinor description naturally extends to complex momenta, which are essential for on-shell recursive techniques, which are based on complex analysis.

\medskip

Before immersing into the actual matter, let us mention those works of other authors that inspired most the present manuscript. Among its spiritual fathers, this work particularly owes to its predecessor~\cite{Conde:2014mdv}, the notes of the lectures given on similar topics by Eduardo Conde, at the ninth edition of the same Modave School, in 2013. Moreover, the co-authored work with Eduardo~\cite{Conde:2016vxs} constitutes a significant part of the material of these notes. Other main inspirational sources for the adopted overall perspective are Weinberg's textbook on Quantum Field Theory~\cite{Weinberg:1995mt}, the more modern book by Henn and Plefka~\cite{Henn:2014yza}, together with the already cited review by Elvang and Huang~\cite{Elvang:2013cua}.

More specifically, for Section~1, where we briefly sketch some preliminary notions from $S$-matrix theory, we refer to Chapter 3 and 4 of Weinberg's book~\cite{Weinberg:1995mt}, to the historical book~\cite{Eden:1966ab}, and to Conde's notes~\cite{Conde:2014mdv}. In Section~2 we review the Lorentz Little Group for massless and massive representations. Then in Section~3 we introduce the spinor-helicity formalism, both for massless and massive momenta, establishing our notation, which is the same as in the useful practical compendium~\cite{Dreiner:2008tw}. In Section~4, essentially based on the already cited work~\cite{Conde:2016vxs}, we show how the constraints coming from Lorentz symmetry completely fix the three-point amplitude in spinor-helicity formalism, for either massless or massive external particles. For Section~5, where we present BCFW recursive techniques and some illustrative examples, we are grateful to the works:~\cite{Elvang:2013cua,Dixon:2013uaa,Conde:2014mdv,Benincasa:2007xk,Brecht:Thesis:2012}.

\vspace{2em}
\noindent\hrulefill
\vspace*{-25mm}
\tableofcontents
\markboth{Introduction}{}
\vspace*{1mm}
\noindent\hrulefill

\pagebreak
%\newpage
\section{Preliminaries on the S-Matrix}	\label{preliminaries}

Let us start with setting some preliminary definitions regarding the $S$-matrix. The probability amplitude of a scattering process is defined \emph{asymptotically}, that is at past and future infinity: we have some initial (interaction-free) particle content, and after a `long' time, during which some interaction happens, we get some other final (interaction-free) particle content. More formally, the probability amplitude of the transitions between some initial state~$|i\rangle$ of the physical Hilbert space at time~$-\infty$ and some final state~$|f\rangle$ at time~$+\infty$ is defined as the expectation value (inner product on the Hilbert space)
\begin{equation}	\label{Smatrix}
S_{fi}=\langle f | i \rangle \ .
\end{equation}
This defines the elements of the $S$-matrix. We should intend the final and initial states as ranging over an orthonormal basis of \emph{multi-particle} states. So, from the completeness relation, it follows that the $S$-matrix has to be \emph{unitary}.

We can subtract from the $S$-matrix the trivial case of transitions with no interaction happening, to get the part that actually contains the interactions:
\begin{equation}
S_{fi}-\delta_{fi} = T_{fi} \ .
\end{equation}

Then, we demand \emph{translational invariance} of the $S$-matrix, namely 
\[
\langle f | i \rangle = \langle e^{-ia_\mu \mathfrak{p}_f^\mu}f | e^{-ia_\mu \mathfrak{p}_i^\mu}i \rangle = e^{-ia_\mu \left(\mathfrak{p}_f^\mu-\mathfrak{p}_i^\mu\right)}\langle f | i \rangle \ ,
\]
where $\mathfrak{p}_i$ represents the sum of incoming four-momenta, and $\mathfrak{p}_f$ the sum of outgoing four-momenta. Of course, for arbitrary $a_\mu$, this identity can be true only if the four-momentum is conserved. So we can extract an overall delta function of momenta and write what we will actually call the amplitude~$M$:
\begin{equation}
T_{fi}=-2\pi i\; \delta(\mathfrak{p}_f-\mathfrak{p}_i)\ M_{fi}\ ,
\end{equation}
with some conventional numerical factor. Then the unitarity condition for $S$ translates into the following condition on $M$:
\begin{equation}	\label{unitarity}
S^\dagger S = \mathbb{I} = S S^\dagger \quad\Rightarrow\quad M-M^\dagger = 2\pi i\: M\,M^\dagger \ .
\end{equation}

Finally, we use the \emph{crossing} symmetry of scattering amplitudes, that is the equivalence of interpreting outgoing particles as incoming antiparticles. In this way we can always treat all the particles as incoming, and replace $M_{fi}$, where we have $m$ incoming and $n-m$ outgoing particles, by $M_n$, where all particles are incoming.

Another crucial property of the $S$-matrix is the \emph{cluster decomposition principle}\label{cluster}. This is just a consequence of the assumption that experiments that are sufficiently distant in space are uncorrelated, which is a mild version of \emph{locality} of the interactions. The consequence is that the total $S$-matrix of distant, uncorrelated scattering processes factorizes into the product of the S-matrices of each of these processes. Then we can restrict to the part of the $S$-matrix where there is an actual exchange of momentum among all the $n$ particles involved (that is, we look at a scattering event in a given, localized experiment), which is called the connected part of the $S$-matrix.

The advantage we will gain from considering the connected part is that it contains no delta function other than the total momentum conservation one, whereas the non-connected part contains additional delta functions corresponding to subsets of particles going through the process without interacting with the rest. On the other hand, the non-connected pieces can be recursively constructed from the lower-point connected one, so the connected amplitudes are really the objects we need to determine in order to have the whole $S$-matrix. Thus, in the following, by $M_n$ we will always implicitly mean the connected component of the amplitude, rather then the whole $n$-point amplitude. For more details about the cluster decomposition principle we refer to standard literature on the subject (for instance chapter 4 of Weinberg's QFT book~\cite{Weinberg:1995mt}).

We have then just sketched the basic properties of $S$-matrix: unitarity, crossing, cluster decomposition (which is related to locality). We have also already imposed translational invariance, but in case of a larger spacetime symmetry group we can impose further constraints on the $S$-matrix. It is precisely what we are going to do, for the four-dimensional Lorentz group. 

In order to do it systematically, we will start reviewing the representation theory of Poincar\'e group, which classifies the different kinds of one-particle states, furnishing a basis for our physical Hilbert space. This is crucial for the definition of $S$-matrix itself, as it is clear from the defining relation~\eqref{Smatrix}.

\section{Poincar\'e representations and the Little Group} \label{PoincareLG}

So, as we have just seen, the scattering amplitude is defined out of the states of the physical Hilbert space, which in turn are classified by the irreducible unitary representations of the symmetry group of spacetime ({\PC} group in our case). Moreover, the states will transform under the symmetry transformations in a determined way, and so the amplitudes will inherit such transformation properties.

So, our preliminary effort is about reviewing the representation theory of Poincar\'e group. We will particularly focus on the Little Group, which will play a central role in the subject of Section~\ref{3pAmpli}, providing the crucial constraints for the three-point amplitude.

The algebra of Poincar\'e group possesses two Casimir operators. A Casimir is an element of the universal enveloping algebra of a given Lie algebra $\mathfrak{g}$, that commutes with all the elements of the algebra $\mathfrak{g}$. Since an irreducible representation has no Casimir beside the identity, the Casimir operators of a group can be used to classify its irreducible representations. The classification of representations of Poincar\'e group in four dimensions is due to Wigner~\cite{10.2307/1968551}, and it is referred to as Wigner classification.\footnote{
	In these notes we will remain in four dimensions, but an analogous classification can be realized for any dimensions bigger than one (see for instance~\cite{Bekaert:2006py}).
}

The Casimir operators of Poincar\'e group are the squared norm of the translation generator (momentum),~$P^2$, and the squared norm of the Pauli-Lubanski operator,~$W^2$, where
\begin{equation}	\label{PLop}
W_\lambda=\frac12\epsilon_{\lambda\mu\nu\rho}M^{\mu\nu}P^\rho \ ,
\end{equation}
with $M^{\mu\nu}$ the generators of the Lorentz algebra.

Since the Casimir operators commute with all other transformations of the group, their eigenstates can be chosen as the physical states: the respective eigenvalues are not affected by any transformation of the group and can be considered as intrinsic properties of the state/particle.

The eigenvalue of the square momentum operator is the squared mass. This divides the Hilbert space into separated classes, whether the mass is zero or different from zero, respectively leading to massless and massive representations of the {\PC} group\footnote{
	Tachyonic momenta (\emph{i.e.} with negative squared mass) constitute a third distinct class, but we discard them as unphysical here. The vacuum ($p\equiv0$) also corresponds to a separate, yet trivial, case.}.

The eigenvalues connected to the Pauli-Lubanski operator, instead, will be related to the helicity/spin of the particle. The Pauli-Lubanski operator generates the Little Group (LG) of a given four-momentum $p$. The LG of $p$ is the stabilizer subgroup of the Lorentz group with respect to~$p$, which is defined as the subgroup of proper orthochronous Lorentz transformations that leave $p$ invariant:
\begin{equation}
\mathrm{LG}_p = \left\{\Lambda_p \in \mathrm{L}_+^\uparrow \:\big/\; \Lambda_p p = p \right\}\ .
\end{equation}

We can thus label our physical states~$\st{p}{a}$ thanks to the eigenvalues of the momentum operator (\emph{i.e.}~$p$) and of Pauli-Lubanski operator (represented for the moment by the generic label~$a$), and decompose the action of any unitary representation~$U$ of a generic Lorentz transformation~$\Lambda$ in the following way
\begin{equation}\label{reps}
U(\Lambda)\st{p}{a} = \sum_{a'}D_{aa'}(\Lambda) \st{\Lambda p}{a'}\ .
\end{equation}
If the considered Lorentz transformation is a LG transformation, then the expression~\eqref{reps} clearly reduces to
\begin{equation}\label{LGreps}
U(\Lambda_p)\st{p}{a} = \sum_{a'}D_{aa'}(\Lambda_p) \st{p}{a'}\ ,
\end{equation}
which does not touch~$p$, but can affect the other labels~$a$, related to the LG.

In the next sections we will see how the Pauli-Lubanski operator actually generates the LG, defining a basis of physical states, and we will explicitely determine the form of the representations~$D_{aa'}$. Since the LG will highlight the fundamentally different nature of massless and massive particle, we will discuss light-like and time-like momenta in distinct sub-sections.

\subsection[LG for massless momenta]{Little Group for massless momenta}

Let us then consider a light-like four-vector~$p$, so meaning~$p^2=0$. There exists a Lorentz transformation $L_p$ that moves $p$ to the special frame
\begin{equation}	\label{kframe}
L_pp\equiv k = \big(E,0,0,E\big)\ .
\end{equation}
Then transformations $\Lambda_k$ of the LG of $k$ would give transformations~$\Lambda_p$ of the LG of~$p$ through the composition~$\Lambda_p={L_p}^{-1}\Lambda_k\,L_p$.

The LG of $k$ can be intuitively identified with the group of isometries {\ISO} in the $p_1\text{\textit{-}}p_2$~plane. Let us verify that the Pauli-Lubanski operator corresponding to~$k$ indeed generates the {\ISO} group. From the definition~\eqref{PLop}, we have
\begin{align*}
&	W_0 = E\,M^{12} \equiv E\,J^3 \ ; \\
&	W_1 = E \left(-M^{23}-M^{02}\right) \equiv -E \left(J^1+K^2\right) \ ; \\
&	W_2 = E \left(-M^{31}+M^{01}\right) \equiv -E \left(J^2-K^1\right) \ ; \\
&	W_3 = -E\,J^3 = -W_0 \ ; 
\end{align*}
where $J^i$ are spatial rotations around the respective axis, and $K^i$ are Lorentzian boosts along the respective direction. The operators $W_1$, $W_2$, $J^3$ verify the algebra (following from Lorentz commutation relations):
\begin{align*}
\big[W_i,W_j\big]=0\ , 	&&	\big[J^3,W_i\big]=i\epsilon_{ij3}\,W_j\ , 	&&	\text{with }\ i,j=1,2 \ ,
\end{align*}
which is indeed the algebra of {\ISO}.

Thus a generic LG transformation for $k$ would act on a massless state~$\st{k}{a}$ as 
\[
e^{-i\alpha W_1}e^{-i\beta W_2}e^{-i\theta J^3}\,\st{k}{a}\ .
\]
The eigenstates of~$W_1,W_2$ turn out to have continuous eigenvalues, which would lead to continuous spin. Even if such possibility constitutes current matter of research~\cite{Schuster:2013vpr,Schuster:2013pxj,Schuster:2013pta,Schuster:2014hca}, it is discarded in standard particle physics (requiring the action of~$W_1,W_2$ to be trivial on the physical states), in favor of a quantized spin. In fact, the eigenstates of~$J^3$ possess discrete eigenvalues, which correspond to the two opposite values of the helicity of a massless particle. This can be seen directly from the definition of helicity, that is the projection of the spin onto the direction of motion; with our special choice of four-momentum $k$ \eqref{kframe}, we have indeed:
\begin{equation}	\label{defHelicity}
H=\frac{\vec{J}\cdot\vec{P}}{|\vec{P}|}	\quad\Rightarrow\quad
	H_k=\frac{\vec{J}\cdot\vec{k}}{|\vec{k}|}=\frac{E\,J^3}{E}=J^3 \ .
\end{equation}

Then we choose the eigenstates of the helicity operator as our physical \emph{massless} particle-states, \emph{i.e.}
\begin{equation}\label{Hstate}
H\st{p}{h}=h\st{p}{h}\ , \quad\text{for }\ p^2=0 \ ,
\end{equation}
where the helicity eigenvalue can take two opposites values~$h=\pm s$, being $s$ the spin of the massless particle. Then, a general little group transformation will act as follows on the state in the special frame~$k$:
\begin{equation}\label{J3st}
U(\Lambda_k)\st{k}{h}=e^{-i\theta J^3} \st{k}{h}=e^{-i\theta h}\st{k}{h}\ ;
\end{equation}
and equivalently for a general light-like four-momenta~$p$:
\begin{equation}
U(\Lambda_p)\st{p}{h}=U(L_p^{-1}\Lambda_kL_p)\st{p}{h}=e^{-i\theta h}\st{p}{h}\ .
\end{equation}
We notice, comparing this expression to the more general~\eqref{LGreps}, that the representations of massless LG are diagonal in the helicity basis, \emph{i.e.}~$D_{hh'}=\delta_{hh'}e^{-i\theta h}$; which is obvious, because the LG is one-dimensional, so it has only one generator and we have chosen as basis the eigenstates of such generator. Yet, this is possible only because the helicity is a Lorentz invariant (Lorentz transformations cannot reverse the direction of motion of a massless particle). It will not be the case for massive representations, as we will see in next section.

\subsection[LG for massive momenta]{Little Group for massive momenta}

As in the massless case, we can bring a generic time-like momentum~$P$, $P^2=m^2$, to a special frame (the rest frame), through a certain Lorentz boost~$L_P$,
\begin{equation}	\label{Kframe}
L_P P\equiv K = \big(m,0,0,0\big)\ ,
\end{equation}
and then we can retrieve the LG transformations for $P$ from those for $K$: $\Lambda_P={L_K}^{-1}\Lambda_KL_K$. From~\eqref{Kframe} we can realize that the LG would be given in this case by the group of three-dimensional spatial rotations~\SO. Again we can derive the generators of the LG from the Pauli-Lubanski operator~\eqref{PLop},
\begin{align*}
&	W_0 = 0 \ ; \\
&	W_i = m\,\epsilon_{ijk0}M^{jk} \equiv -m\,J^i \ . 
\end{align*}
Of course, the generators of spatial rotations~$J^i$ by definition reconstruct the algebra of \SO, which is equivalent to the algebra of~$\SU$, \emph{i.e.}
\begin{equation}
\big[J^i,J^j\big]=i\epsilon^{ijk}J^k .
\end{equation}
In the rest frame, the total angular momentum, which is in general the sum of orbital and intrinsic angular momentum, $\vec{J}=\vec{L}+\vec{S}$, is given just by the intrinsic angular momentum, the spin $\vec{S}$. The Casimir is thus $W^2=\vec{J\:}^2=\vec{S\:}^2$, and we choose the particle-states to be eigenstates of the Casimir, with eigenvalues $s\,(s+1)$, where $s$ defines the spin of the massive particle. Yet, as we know from our quantum mechanics courses, this is not sufficient to define a basis of the Hilbert space: we need to choose the component of the spin along one direction (which we will call $J_0$), and the corresponding eigenstates with eigenvalue~$\sigma$ will give the complete basis. Then we can label our massive states by $s$ and $\sigma$, and write
\begin{align}
\vec{J\,}^2\st{P}{s,\sigma} 	&= s\left(s+1\right)\st{P}{s,\sigma}\ ; \\
J^0\st{P}{s,\sigma} 			&= \sigma\,\st{P}{s,\sigma}\ , \label{J0st}\\
J^\pm\st{P}{s,\sigma} 			&= \sigma^\pm\,\st{P}{s,\sigma\pm 1}\ ; \label{Jpmst}
\end{align}
where $\sigma^\pm\equiv\sqrt{(s\mp\sigma)(s+1\pm\sigma)}$, and the generators $J_\pm$ are defined in the standard way in order to satisfy the $\SU$ commutation relations
\begin{equation}\label{SU2alg}
\big[J^+,J^-\big]=2J^0 \ ,\qquad\qquad \big[J^0,J^{\pm}\big]=\pm J^{\pm} \ .
\end{equation}
As announced in previous section, we notice that, since the spin projection~$\sigma$, contrarily to the helicity, is not a Lorentz invariant (it is indeed shifted by transformations generated by $J_+$ and $J_-$), the representations of massive LG transformations will not be diagonal in the eigenstates of $J_0$: 
\begin{equation}	\label{massrep}
U(\Lambda_P)\st{P}{s,\sigma} = \sum_{\sigma'}D_{\sigma\sigma'}(\Lambda_P)\st{P}{s,\sigma'}\ .
\end{equation}
This is a crucial difference that we will have to take into account when we will try to extract constraints for the amplitudes from the massive LG.

\subsection*{Commentary: the LG transformations and the amplitude}

Before moving on, we want to make more explicit the link between LG transformations and the amplitude, which is the physical object we are interested in. The fact is that, since the amplitude is made out of a direct product of in-going (out-going) states, it will inherit the transformation properties of the states under Lorentz transformations. Referring to expression~\eqref{reps}, we can write
\begin{equation}\label{PoincareM}
M_n\big(\{p_i\};\{a_i\}\big) \;\overset{\Lambda}{\longrightarrow}\; \bigg(\sum_{a'_j}D_{a_ja'_j}(\Lambda,p_j)\bigg)\,M_n\big(\{\Lambda p_i\};\{a'_i\}\big) \ ,
\end{equation}
where the Lorentz transformation is acting on the $j$-th state inside the $n$-point amplitude~$M_n$ with $n$ external momenta $p_i$. In particular, for infinitesimal LG transformations it reads
\begin{equation}\label{J3M}
H_jM_n\big(\{p_i\};\{a_i\}\big) = h_j\,M_n\big(\{p_i\};\{a_i\}\big) \ ,
\end{equation}
for $p_j$ massless, from eq.~\eqref{J3st}, and
\begin{equation}	\label{J0pmM}
\begin{aligned}
	J^0_jM_n\big(\{p_i\};\{a_i\}\big) &= 	\sigma_j\,M_n\big(\{p_i\};\ldots, \sigma_j,\ldots\big) \ , \\
		J^\pm_jM_n\big(\{p_i\};\{a_i\}\big) &= 	\sigma^\pm_j\,M_n\big(\{p_i\};\ldots, \sigma_j\pm 1,\ldots\big) \ ,
\end{aligned}
\end{equation}
for $p_j$ massive, from eq.s~\eqref{J0st} and~\eqref{Jpmst}.

The fact that the amplitude has to transform in a proper way under LG transformations yields strong constraints on the general form of the amplitude. Yet it is not immediate to extract such constraints from equations~(\ref{J3M}-\ref{J0pmM}), we need to write them in a more explicit way. For such purpose, the spinor-helicity formalism furnishes a language to translate LG equations in a ready-to-use and effective form, in terms of simple linear differential equations.

Hence in next section we will introduce and review the spinor-helicity formalism, before making large use of it in the rest of our discussion.

\section{Spinor-helicity formalism}

The spinor-helicity formalism is an ubiquitous ingredient of the recent \emph{vague} of successes in computing scattering amplitudes with on-shell methods. Yet, it does not contain anything magic. It is just a language which translates null-norm four-vectors, transforming under the $(\frac12,\frac12)$ representation, into a pair of Weyl bi-spinors, transforming under the $(\frac12,0)$ and $(0,\frac12)$ representations. This has the advantage of implementing by construction the massless on-shell condition, as we will see. 

The essential fact behind all this is that the (proper ortochronous) Lorentz group~$\mathrm{L}(\mathbb{R})$ is homomorphic to $\SL$. This two-to-one correspondence can be more naturally understood and explicitly constructed for the complexification of the Lorentz group, \emph{i.e.}~$\mathrm{L}(\mathbb{C})$.\footnote{
	This constitutes a first reason to consider complex momenta throughout the discussion of these notes.}
Using the Pauli matrices plus the identity, $\sigma^{\mu}=\big(\mathbb{I},\vec{\sigma}\big)$, we can associate a complex two-by-two matrix to any four-vector:
\begin{equation}\label{4to2x2}
\begin{array}{crclc}
\mathbb{C}^4 	&\quad\quad&\longrightarrow &\quad\quad& M_2(\mathbb{C})\\
p_{\mu}=(p_0,p_1,p_2,p_3) 	&&\longmapsto&& \sigma^{\mu}_{a\dot{a}}p_{\mu}=\left(\begin{array}{cc} p_0+p_3 & p_1-ip_2\\p_1+ip_2 & p_0-p_3\end{array}\right)=p_{a\dot{a}}\;\;.
\end{array}
\end{equation}
In this a way, a complex Lorentz transformation acting on a complex four-vector corresponds to the action of two $\SL$ transformations conjugately acting on the complex two-by-two matrix.
\begin{equation}\label{homomap}
{\renewcommand{\arraystretch}{1.7}
\begin{array}{crclc}
\mathrm{L}(\mathbb{C}) & 
	\quad\quad&\longrightarrow&\quad\quad& 
		\SL\times\SL\\
\Lambda: p_{\mu}\mapsto \Lambda^{\phantom{\mu}\nu}_{\mu} k_{\nu} &&
	\longmapsto &&
		\zeta(\Lambda),\xi(\Lambda):p_{a\dot{a}} \mapsto \zeta^{\phantom{a}b}_a\,p_{b\dot{b}}\,{\xi}^{\dot{b}}_{\phantom{b}\dot{a}}
\end{array}}
\end{equation}
Demanding the transformed four-vector to match the transformed two-by-two matrix, we derive the defining map between $\Lambda$ and $\zeta(\Lambda),\xi(\Lambda)$:
\begin{equation}\label{L-SL}
\left.\begin{aligned}
&
	p'_{a\dot{a}}=\zeta(\Lambda)^{\phantom{a}b}_a p_{b\dot{b}}{\xi(\Lambda)}^{\dot{b}}_{\phantom{a}\dot{a}} = \zeta(\Lambda)^{\phantom{a}b}_a\sigma^{\nu}_{b\dot{b}}{\xi(\Lambda)}^{\dot{b}}_{\phantom{a}\dot{a}}p_\nu \phantom{\frac{}{|}} \\
&
	p'_{a\dot{a}}= \sigma^{\mu}_{a\dot{a}}p'_\mu = \sigma^{\mu}_{a\dot{a}}\Lambda^{\phantom{\mu}\nu}_{\mu}p_{\nu}
\end{aligned}\right\} \;\Longrightarrow\;
	\zeta(\Lambda)^{\phantom{a}b}_a \sigma^{\mu}_{b\dot{b}}{\xi(\Lambda)}^{\dot{b}}_{\phantom{a}\dot{a}} = \sigma^{\nu}_{a\dot{a}}\Lambda^{\phantom{\nu}\mu}_{\nu}
\end{equation}
The fact that the transformation matrices~$\zeta,\xi$ have to belong to $\SL$ descends from the defining property of the Lorentz group, the conservation of the norm: $(\Lambda p)^2=p^2$. In the two-by-two matrices language the norm is given by the determinant of the matrix:
\begin{equation}
\det\!|p|=p_0^2-\vec{p\,}^2=p^2\ .
\end{equation}
Then
\begin{equation}\label{det1}
\det\!|p|=\det\!|\zeta p\xi|=\det\!|\zeta|\det\!|p|\det\!|\xi| \quad\Leftrightarrow\quad \det\!|\zeta|\det\!|\xi|=1 \ .
\end{equation}
There is a redundancy in such defining property, since we can always rescale $\zeta$ and $\xi$ in the following way,
\begin{equation*}
\begin{aligned}
&	\zeta \longrightarrow C^{-1} \zeta	\\
&	\xi \longrightarrow C\, \xi
\end{aligned}\qquad \text{with }\ C\in\mathbb{C}\ ,
\end{equation*}
without spoiling the condition~\eqref{det1}. Then, if we take $C=\det\!|\zeta|$, the redefined $\zeta$ gets unit determinant, and so $\xi$ as well must have unit determinant, in order to satisfy~\eqref{det1}:
\begin{equation}
\det\!|\zeta|=1=\det\!|\xi| \quad\Longrightarrow\quad \zeta,\xi \in \SL \ .
\end{equation}

We have thus constructed the homomorphism between $\mathrm{L}(\CC)$ and $\SL\times\SL$. It is a homomorphism, and not an isomorphism, because there is still a leftover redundancy in sending $\zeta,\xi$ into $-\zeta,-\xi$, so that we have a two-to-one map. The quotient of $\SL\times\SL$ by $\mathbb{Z}_2$ is then giving a one-to-one map, \emph{i.e.} an isomorphism.

If we want to recover the real-valued case, we have to impose a reality condition $p_\mu^*=p_\mu$, which yields in turns $p_{a\dot{a}}^\dagger=p_{a\dot{a}}$, as a straightforward consequence of the hermiticity of Pauli matrices. Imposing hence this reality condition on the transformed momentum-matrices, we get
\begin{equation}
\zeta p \xi = \left(\zeta p \xi\right)^\dagger = \xi^\dagger p \zeta^\dagger \quad\Leftrightarrow\quad \xi\equiv\zeta^\dagger \ ,
\end{equation}
\emph{i.e.}: the ``right-handed'' transformation must be the conjugate of the ``left-handed'' one. So we have to take the diagonal $\SL$ in the product of~\eqref{homomap} in order to get the homomorphism with the (real) proper orthochronous Lorentz group, $\mathrm{L}(\RR)$.

Finally, we can specialize to massless momenta and define the spinor-helicity formalism. A null four-vector translates in a two-by-two matrix with null determinant. A complex two-by-two matrix with null determinant can be always expressed as the direct product of a pair of complex two-dimensional vectors:
\begin{equation}	\label{plambda}
\det\!|p|=0 \quad\Rightarrow\quad p_{a\dot{a}}=\lambda_a\otimes\tilde{\lambda}_{\dot{a}}=
	\left(\!	\begin{array}{cc}
				\lambda_1\tilde{\lambda}_1 	& \lambda_1\tilde{\lambda}_2 \\
				\lambda_2\tilde{\lambda}_1 	& \lambda_2\tilde{\lambda}_2
				\end{array}\!\right)\ .
\end{equation}
In the following we will keep the tensor product implicit, using the lighter notation: $p=\lambda\tilde{\lambda}$.

You can notice that the advantage of writing a null momentum in such a way is that the on-shell condition~$\det\!|\lambda\tilde{\lambda}|=0$ is built-in, and any expression written in this formalism would be automatically on-shell, with no need to enforce this condition by hand.

We define now a bunch of conventions and shorthand notations for spinor products, which we will extensively use in the rest of these notes. First of all, we can take contractions of two bi-spinors, which are $\SL$ invariant in the same way as scalar products of four-vectors are Lorentz invariant:
\begin{align}
\ket{\lambda}{\mu} &\equiv 
	\lambda^a\mu_a=\epsilon^{ab}\lambda_{b}\mu_{a} \ , 	\label{ket}\\
\bra{\tilde{\lambda}}{\tilde{\mu}} &\equiv
	\lambda_{\dot{a}}\mu^{\dot{a}}=\epsilon^{\dot{a}\dot{b}}\tilde{\lambda}_{\dot{a}}\tilde{\mu}_{\dot{b}} \ ,	\label{bra}
\end{align}
with $\epsilon^{12}=1=\epsilon^{\dot{1}\dot{2}}$, $\epsilon_{12}=-1=\epsilon_{\dot{1}\dot{2}}$, and $\epsilon^{ac}\epsilon_{cb}=\delta^{a}_{\phantom{a}b}$. These inner products are obviously anti-symmetric and, in particular, their vanishing implies that the two spinors are proportional. The Minkowski scalar product translates into a product of two-by-two matrices (non necessarily with null determinant), as follows:
\begin{equation}
2p\cdot{q}=\epsilon^{ab}\epsilon^{\dot{a}\dot{b}}\, p_{a\dot{a}} q_{b\dot{b}} \ .
\end{equation}
In the case of light-like momenta, we can replace the matrices by a pair of spinors, obtaining
\begin{equation}\label{pipj}
2\,p_{i} \cdot p_{j} = \ket{\lambda_{i}}{\lambda_{j}}\bra{\tilde\lambda_{j}}{\tilde\lambda_{i}}\equiv\ket{i}{j}\bra{j}{i}\ ,
\end{equation}
where in the last passage we have introduced a shorthand notation that we will largely use throughout these notes. Another shorthand notation that we will adopt is the following one, for the product of a light-like vector $k=\kappa\tilde{\kappa}$ with a generic four-momentum $p$:
\begin{equation}	\label{Pgen}
2k\cdot{p}=\epsilon^{ab}\epsilon^{\dot{a}\dot{b}}\,\kappa_a\tilde{\kappa}_{\dot{a}}\, p_{b\dot{b}}\equiv\Pgen{\kappa}{p}{\tilde{\kappa}} \ ,
\end{equation}
which naturally reduces to something of the kind of~\eqref{pipj} in the particular case where also $p$ is light-like.

\subsection{Spinor-helicity formalism for massive particles}

The main power of the spinor-helicity description for null momenta is the automatic implementation of the on-shell condition, as we have just seen. If we want to apply this formalism to massive momenta, we loose such advantage, since we do not know how to enforce the massive on-shell condition~$\det\!|p|=m^2$ by construction. Nevertheless, as we will see in next section, a second advantage of spinor-helicity formalism is the simple and effective form that LG operators take in this language. Such simplicity and effectiveness is realized both for massless and massive particles. Moreover, if we want to consider amplitudes where massless and massive particles are involved at the same time, it is sensible to treat them on the same footing.

There are two standard ways of expressing a time-like momentum~$P$ in terms of bi-spinors, as discussed by Dittmaier~\cite{Dittmaier:1998nn}. We choose the strategy of representing $P$ as the sum of two null momenta, $p$ and $q$:
\begin{equation}\label{Paa}
P_\mu=p_\mu+q_\mu \quad\Rightarrow\quad P_{a\dot{a}}=\lambda_a\tilde\lambda_{\dot{a}}+\mu_a\tilde\mu_{\dot{a}} \ ,\qquad
2p\cdot{q}=\ket{\lambda}{\mu}\bra{\tilde\mu}{\tilde\lambda}=m^2 \ .
\end{equation}
Notice that there are many ways of decomposing a time-like vector in terms of two light-like vectors. This description introduces thus a redundancy, which is not physical and will need to be removed at a certain point. We should not confuse this non-uniqueness with the one that yet we have for massless momenta described in terms of bi-spinors~\eqref{plambda}: in this latter case we can act with $\SL$-transformations on $\lambda$ and $\tilde{\lambda}$ such that the final two-by-two matrix is left unchanged. But these are nothing else that the physical LG transformations, which, as we will see later on, will be precious for constraining the amplitude. On the contrary, the physical massive LG invariance will mix up with the non-physical redundancy introduced by our ambiguous decomposition~\eqref{Paa}. It will be crucial in section~\eqref{mass3ampl} to distinguish the two of them, and to remove the latter.

We have thus briefly seen how to apply the spinor-helicity formalism to massive particles in a very simple way, which yet yields some disadvantages. Now we will see how to write the LG equations in this formalism, bringing out the advantages of our choice.

\section{Constraining the amplitude: the LG equations}	\label{3pAmpli}

The aim of this section is to translate equations~\eqref{J3M} and~\eqref{J0pmM} in spinor language, and use them to constrain the amplitude. In this context, the asymptotic states, instead of being labeled in terms of massless or massive four-momenta, will be labeled by pairs of bi-spinors, $\{\lambda_i,\tilde{\lambda}_i\}$. 

Since the LG for massless and massive momenta is fundamentally different, we will treat the two cases separately. The derived LG equations yield constraints on the dependency of the n-point amplitude on its kinematic variables (spinor products). For the 3-point amplitude, such constraints will be enough to completely determine the form of the amplitude, both in the massless and in the massive case.

\subsection{Massless LG equations}

We have seen that the massless LG corresponds to rotations around the direction of motion. For instance, in the special frame $k_\mu=(E,0,0,E)$ it is given by rotations around the third axis. These are a subgroup of (real) Lorentz transformations parametrized by an angle $\theta$, acting on a four-momentum $p$ as follows:
\begin{equation*}
R_3(\theta)p =
\left(\begin{array}{cccc}
	1&0&0&0\\
	0&\cos(\theta)&-\sin(\theta)&0\\
	0&\sin(\theta)&\cos(\theta)&0\\
	0&0&0&1
	\end{array}\right)
\left(\begin{array}{c}
		p_0\\
		p_1\\
		p_2\\
		p_3
		\end{array}\right)\ .
\end{equation*}
How this transformation translates into $\SL$ transformations acting on bi-spinors? The answer can be worked out straightforwardly from the relation~\eqref{L-SL}, and is given by
\begin{equation}	\label{zetaR3}
{\zeta_{R_3}(\theta)}^{\phantom{a}b}_a\lambda_b = \pm e^{-i\frac{\theta}{2}\sigma^3}\lambda=\pm
\left(\begin{array}{cc}
	e^{-i\theta/2} & 0\\
	0 & e^{i\theta/2}
	\end{array}\right)
\left(\begin{array}{c}
		\lambda_1\\
		\lambda_2
		\end{array}\right)\ ,
\end{equation}
where the sign ambiguity corresponds to the $\mathbb{Z}_2$ ambiguity in the homomorphism between Lorentz and $\SL$. The special frame $k$ in spinor language is given by
\begin{equation*}
k_{a\dot{a}}=\kappa_a\kappa^\dagger_{\dot{a}}=
	\left(\begin{array}{cc}2E&0\\0&0\end{array}\right)\ , \qquad
		\text{with }\ \kappa_a=\left(\begin{array}{c}\sqrt{2E}\\0\end{array}\right)\ .
\end{equation*}
Hence, whereas the momentum-matrix is trivially invariant under LG transformations, \emph{i.e.} \mbox{$\zeta_{R_3}k\zeta^\dagger_{R_3}=k$}, as it should be, it is not the case for each bi-spinor by itself:
\begin{equation*}
\zeta_{R_3}\kappa=e^{-i\frac{\theta}{2}}\kappa\ , \quad
	\kappa^\dagger\zeta^\dagger_{R_3}=e^{+i\frac{\theta}{2}}\kappa^\dagger \ .
\end{equation*}
So the LG is acting non-trivially on the spinors (yielding a phase), and actually these transformations, which we have here derived for the special frame~$k$, are the same for any generic frame (see the appendix A of~\cite{Conde:2016vxs} for more details), namely
\begin{equation}\label{LGscaling}
\lambda \;\longrightarrow\; e^{-i\frac{\theta}{2}}\,\lambda \ , \qquad
	\tilde{\lambda} \;\longrightarrow\; e^{+i\frac{\theta}{2}}\tilde{\lambda}\ .
\end{equation}
In any case, however, the momentum matrix~$p$ is conserved under such transformations, whereas the momentum-spinors are scaling in a precise way, and so the amplitude scales accordingly.

The differential operator that generates the infinitesimal version of transformations~\eqref{LGscaling} is
\begin{equation} \label{Hop}
H=-\frac{1}{2}\left(\lambda^a\bfrac{\textstyle\partial\phantom{\lambda^a}}{\partial\lambda^a}-\tilde\lambda_{\dot{a}}\bfrac{\partial\phantom{\lambda^a}}{\partial\tilde\lambda_{\dot{a}}}\right) \equiv 
	-\frac{1}{2}\left(\lambda\bfrac{\partial\phantom{\lambda}}{\partial\lambda}-\tilde\lambda\bfrac{\partial\phantom{\lambda}}{\partial\tilde\lambda}\right) \ ,
\end{equation}
so that the helicity equation~\eqref{Hstate} becomes
\begin{equation}\label{Hp}
\bigg(\lambda\bfrac{\partial\phantom{\lambda}}{\partial\lambda}-\tilde\lambda\bfrac{\partial\phantom{\lambda}}{\partial\tilde\lambda}\bigg)\st{\lambda,\tilde{\lambda}}{h}=-2h\:\st{\lambda,\tilde{\lambda}}{h}\ ,
\end{equation}
for the massless one-particle state, and consequently for the $n$-point amplitude,
\begin{equation}\label{HMn}
\bigg(\lambda_j\bfrac{\partial\phantom{\lambda}}{\partial\lambda_j}-\tilde\lambda_j\bfrac{\partial\phantom{\lambda}}{\partial\tilde\lambda_j}\bigg)\,M_n\big(\{\lambda_i,\tilde{\lambda}_i\},\{a_i\}\big) =-2h_j\:M_n\big(\{\lambda_i,\tilde{\lambda}_i\},\{a_i\}\big) \ ,
\end{equation}
when the $j$-th particle is massless, from eq.~\eqref{J3M}. This equation constitutes a sort of Ward identity for the $n$-point amplitude, and it is as powerful as to completely constrain the form of the amplitude for the lowest-point case, \emph{i.e. }$n=3$, as it was first derived by Benincasa and Cachazo~\cite{Benincasa:2007xk}, and as we will see immediately.

\subsection{The massless three-point amplitude}\label{null3ampl}

The three-point amplitude for three massless particles has to be zero for real external momenta: a massless particle cannot decay into two other massless particles, except for aligned momenta and helicities summing to zero. This can be seen just by applying momentum conservation to the case of three light-like four-vectors, as it is shown in~\cite{Adler:2016peh}. Momentum conservation can be satisfied only if the spatial momenta are aligned. Then, being the helicity the projection of the spin along the direction of motion, the conservation of the spin imposes $h_1+h_2+h_3=0$\footnote{
	This turns out to be a somehow sick case (see the comments on page~\pageref{sick}), except for the case of three massless scalars, where yet the 3-point amplitude is just a constant, the cubic scalar coupling.}.

If we allow for complex momenta, instead, we can have non-trivial three-point amplitudes for arbitrary values of the helicities. The advantage of considering complex momenta and so having non-zero massless three-point amplitudes will be clear in section~\ref{BCFW}, when we will use the power of complex analysis to glue together two three-point amplitudes to obtain a four-point one. Sending afterwards the complex momenta to real ones, we will get a non-zero final result, which is the actual physical four-point amplitude with real-valued external momenta.

So we derive now the most general Poincar\'e-invariant three-point function with complex massless external momenta, keeping in mind that it will constitute the fundamental brick to construct higher-point amplitudes. To do that, we just consider three times the LG equation~\eqref{HMn} for~$n=3$:
\begin{equation}\label{HM3}
\bigg(\lambda_j\bfrac{\partial\phantom{\lambda}}{\partial\lambda_j}-\tilde\lambda_j\bfrac{\partial\phantom{\lambda}}{\partial\tilde\lambda_j}\bigg)\,M_3^{h_1,h_2,h_3}\big(\{\lambda_i,\tilde{\lambda}_i\}\big) =-2h_j\:M_3^{h_1,h_2,h_3}\big(\{\lambda_i,\tilde{\lambda}_i\}\big) \ , \quad\text{with }\ j=1,2,3 \ .
\end{equation}
The amplitude will depend only on Lorentz invariants, $\SL$ invariants in our case, that is the scalar products of spinors~(\ref{ket}-\ref{bra}). Then it is convenient to change to those variables:
\begin{equation}
\begin{array}{lll}
\vphantom{\frac{}{\big|}} 
	x_1=\ket23 \ , \quad	& x_2=\ket31 \ , \quad	& x_3=\ket12 \ ;	\\
		y_1=\bra32 \ , \quad	& y_2=\bra13 \ , \quad	& y_3=\bra21 \ ;
\end{array}
\end{equation}
where we have used the shorthand notation~\eqref{pipj}. If we use the chain rule, \emph{i.e.}
\[
\lambda_1\bfrac{\partial\phantom{\lambda}}{\partial\lambda_1}=x_2\bfrac{\partial\phantom{x}}{\partial x_2}+x_3\bfrac{\partial\phantom{x}}{\partial x_3}\ ,
\]
and so on, we can recast the three equations~\eqref{HM3} in the following way (omitting the subscript~3 on $M$ from now on):
\begin{equation}\label{eqsm0M3}
\begin{aligned}
\left(x_1\partial_1-y_1\tilde{\partial}_1\right) M^{\{h_{i}\}}\!\big(\{x_i,y_i\}\big) 
& 	= \big(h_1-h_2-h_3\big)\, M^{\{h_{i}\}}\!\big(\{x_i,y_i\}\big) \ , \\
\left(x_2\partial_2-y_2\tilde{\partial}_2\right) M^{\{h_{i}\}}\!\big(\{x_i,y_i\}\big) 
& 	= \big(h_2-h_3-h_1\big)\, M^{\{h_{i}\}}\!\big(\{x_i,y_i\}\big) \ , \\
\left(x_3\partial_3-y_3\tilde{\partial}_3\right) M^{\{h_{i}\}}\!\big(\{x_i,y_i\}\big) 
& 	= \big(h_3-h_1-h_2\big)\, M^{\{h_{i}\}}\!\big(\{x_i,y_i\}\big) \ ;
\end{aligned} 
\end{equation}
where we have used the shorthand notations $\partial_i$ and $\tilde{\partial}$, for partial derivatives with respect to $x_i$ and $y_i$ respectively.

The most general solution\footnote{
	This kind of solutions for partial differential equations are standardly obtained through the \href{https://en.wikipedia.org/wiki/Method_of_characteristics}{method of characteristics}.
} for this system of equations is
\begin{equation}\label{M3nom}
\begin{aligned}
M^{\{h_{i}\}}\!\big(\{x_i,y_i\}\big) &= 
	x_1^{h_1-h_2-h_3}\, x_2^{h_2-h_3-h_1}\, x_3^{h_3-h_1-h_2}\; f\big(x_1y_1,\,x_2y_2,\,x_3y_3\big) \\
&=	y_1^{h_2+h_3-h_1}\, y_2^{h_3+h_1-h_2}\, y_3^{h_1+h_2-h_3}\; \tilde{f}\big(x_1y_1,\,x_2y_2,\,x_3y_3\big) \\
\end{aligned}\quad ,
\end{equation}
where we have a pre-factor encoding the proper LG scaling, and an undetermined function depending on the combinations $x_iy_i$, which precisely vanishes under the action of the differential operators $x_i\partial_i-y_i\tilde{\partial}_i$. We have written the solution in two different ways, just to make manifest that there are many equivalent ways to write the pre-factor, and in order to retrieve the actual form of the amplitude we need to explicitly determine the function~$f$ (or~$\tilde{f}$).

Actually, we still have to impose momentum conservation, which yields relations among our six variables. We will discuss these relations for generic~$n$ external momenta in section~\ref{mass3ampl}, but for three massless momenta the analysis is straightforward. We obtain for instance
\[
0=p_1^2=(-p_2-p_3)^2=2p_2\cdot p_3=\ket{2}{3}\bra{3}{2}= x_1y_1 \ .
\]
So we have to choose $x_1=0$, or $y_1=0$. Let us try with $x_1=\ket{2}{3}=0$. This means that $\lambda_2$ and $\lambda_3$ are proportional (aligned in the two-dimensional vector space of the spinors). \label{lindep}In addition, in a two-dimensional vector space three vectors cannot be linearly independent, so $\lambda_1=\alpha\lambda_2+\beta\lambda_3$. Therefore all the `non-tilded' spinors have to be proportional: $\lambda_1\propto\lambda_2\propto\lambda_3$. Of course, if we had started with $y_1=0$, we would have obtained that all the `tilded' spinors should be proportional. Thus momentum conservation requires that all~$x_i=0$, or all~$y_i=0$.

So there are only two forms for the amplitude to be finite and not trivially zero\footnote{
	We remind that we are considering the connected part of the amplitude, so that we know that it does not contain any delta functions.
}, which are
\begin{align}
\label{M3H}
&M^{\{h_i\}}_H=g_H\; x_1^{h_1-h_2-h_3}\, x_2^{h_2-h_3-h_1}\, x_3^{h_3-h_1-h_2} \ , \qquad 
	\text{when }\ y_i=0\;\ \forall\,i\ ,  \vphantom{\frac{}{\big|}}\\
\label{M3A}
&M^{\{h_i\}}_A=g_A\; y_1^{h_2+h_3-h_1}\, y_2^{h_3+h_1-h_2}\, y_3^{h_1+h_2-h_3} \ , \qquad 
	\text{when }\ x_i=0\;\ \forall\, i\ ,
\end{align}
respectively corresponding to the following values of the function~$f$:
\begin{equation}\label{2fs}
f = g \ , 	\quad\textrm{and}\quad 
	f = \tilde{g}\, \left(x_1y_1\right)^{h_2+h_3-h_1} \left(x_2y_2\right)^{h_3+h_1-h_2} \left(x_3y_3\right)^{h_1+h_2-h_3} \ .
\end{equation}

It is eventually possible, by imposing an additional physical requirement, to determine which of the two forms of the amplitude is the correct one, depending on the values of the helicities. We remind the reader our initial remark of this section: for real external momenta the massless three-point function is trivially zero (except for specific values of the helicities, \emph{s.t.} $h_1+h_2+h_3=0$).\label{sick} The reality conditions read $y_i=x_i^*$, implying that $x_i=0=y_i$ for every~$i$. Then the amplitude~\eqref{M3H} vanishes (and does not explode) for $h_1\!+\!h_2\!+\!h_3<\!0$, whereas the amplitude~\eqref{M3A} vanishes (and does not explode) for $h_1\!+\!h_2\!+\!h_3>\!0$. The choice remains ambiguous precisely for the case $h_1+h_2+h_3=0$, where actually the amplitude can be not vanishing. But except the trivial case of three scalars, where the amplitude is just a constant, \emph{i.e.}~the cubic coupling, there is no known interaction that yields a three-particle process with $h_1+h_2+h_3=0$, and actually it can be shown that, for theories where we can construct higher-point amplitudes out of the three-point ones, such interactions are ruled out~\cite{McGady:2013sga}.

Thus the (complex-valued) massless three-point amplitude is fixed by Poincar\'e invariance up to a constant, the coupling constant~$g_H$ or~$g_A$, as it was first derived by Benincasa and Cachazo~\cite{Benincasa:2007xk}. We stress that the results~\eqref{M3H} and~\eqref{M3A} are \emph{non-perturbative}, since they rely only on Poincar\'e invariance plus the requirement that the amplitude be non-singular. This last requirement applies to the full non-perturbative physical amplitude, as well as tree-level amplitudes. In a perturbative expansions, as we all know, intermediate steps are typically non-finite. Such kinds of `partial' amplitudes, as long as they obey Lorentz invariance, should still obey the most general form~\eqref{M3nom}\footnote{
	The general form~\eqref{M3nom} does not appear in the original paper~\cite{Benincasa:2007xk}, but has been instead proposed in~\cite{Conde:2016vxs}, where an example of loop divergent amplitude, matching~\eqref{M3nom} rather than~(\ref{M3H}-\ref{M3A}), is given as well.
}. 

In next section we will extend the successful strategy, that has allowed us to determine the massless three-point amplitude, to the case where one, two, or three massive particles participate in the scattering. The derivation will be less straightforward, but just as successful.

\subsection{Massive LG equations}

As we have seen in section~\ref{PoincareLG}, an $n$-point amplitude that involves a massive particle has to obey the massive LG equations~\eqref{J0pmM} for the corresponding leg. The generators $J_0$, $J_+$, $J_-$ respect the three-dimensional $\SU$ algebra~\eqref{SU2alg}. We want to turn these equations in spinor-helicity language, that is spinor differential operators acting on the amplitude. As in previous section we can consider the action of a LG transformation on a four-vector, and from that infer the corresponding transformations on the spinors.

We recall our description of a massive momentum-matrix in spinor-helicity formalism~\eqref{Paa}, and we go to the rest frame~$K_\mu=(m,0,0,0)$, where we have
\begin{equation*}
K_{a\dot{a}}= \left(
	\begin{array}{cc}
	m	& 0		\\
	0	& m
	\end{array}\right) = \lambda_a\tilde\lambda_{\dot{a}}+\mu_a\tilde\mu_{\dot{a}}= \left(
		\renewcommand{\arraystretch}{1.3}\begin{array}{cc}    
		\lambda_1\tilde{\lambda}_1+\mu_1\tilde{\mu}_1 &\;\;	
			\lambda_1\tilde{\lambda}_2+\mu_1\tilde{\mu}_2 \\
		\lambda_2\tilde{\lambda}_1+\mu_2\tilde{\mu}_1 &\;\; 
			\lambda_2\tilde{\lambda}_2+\mu_2\tilde{\mu}_2
		\end{array}\right)\ ,
\end{equation*}
with the following conditions on the spinor components:
\begin{eqnarray}	\label{Kconditions}
\lambda_{1}\tilde{\lambda}_{2}+\mu_{1}\tilde{\mu}_{2} & = 0 = & \lambda_{2}\tilde{\lambda}_{1}+\mu_{2}\tilde{\mu}_{1}\ , \nn
\lambda_{1}\tilde{\lambda}_{1}-\mu_{2}\tilde{\mu}_{2} & = 0 = & \mu_{1}\tilde{\mu}_{1}-\lambda_{2}\tilde{\lambda}_{2}\ , \\
\lambda_{1}\tilde{\lambda}_{1}+\lambda_{2}\tilde{\lambda}_{2} & = m = & \mu_{1}\tilde{\mu}_{1}+\mu_{2}\tilde{\mu}_{2}\ . \nonumber
\end{eqnarray}
The LG is given by \SO, the three-dimensional spatial rotations. If $R$ is a generic element of \SO, then the corresponding $\SL$ matrices acting on spinors, $\zeta_{R},\,\zeta_R^\dagger$, turn out to be elements of $\SU$ subgroup. This can be easily checked taking rotations around specific reference axis, like the one in~\eqref{zetaR3}, and composing them to obtain a generic rotation, for instance by the standard parametrization through Euler angles. Then we take
\begin{equation}	\label{zetaR}
\zeta_{R}=\left(\begin{array}{cc} a & b \\ -b^{*} & a^{*} \end{array}\right)\ ,\quad \text{with }\ |a|^2+|b|^2=1 \ ,
\end{equation}
and we check that indeed the correct LG transformation property for $K_{a\dot{a}}$ is fulfilled:
\begin{equation}
{\zeta_{R}}^{\phantom{a}b}_a K_{b\dot{b}} {\zeta^\dagger_{R}}^{\dot{b}}_{\phantom{a}\dot{a}} =
	{\zeta_{R}}^{\phantom{a}b}_a\lambda_b\tilde\lambda_{\dot{b}}{\zeta^\dagger_{R}}^{\dot{b}}_{\phantom{a}\dot{a}} + {\zeta_{R}}^{\phantom{a}b}_a\mu_b\tilde\mu_{\dot{b}}{\zeta^\dagger_{R}}^{\dot{b}}_{\phantom{a}\dot{a}} = K_{a\dot{a}} \ .
\end{equation}
The last identity holds in virtue of relations~\eqref{Kconditions}, which are specific to this frame. We also remark that the total momentum is invariant under these $\SU$ transformations, whereas $\lambda_a\tilde{\lambda}_{\dot{a}}$ and $\mu_a\tilde{\mu}_{\dot{a}}$ are not separately invariant. Of course we can consider a generic boosted frame $P=L_P^{-1}K$, and after having found the proper formulation of boosts in terms of $\SL$ matrices, we would obtain the LG transformations for a generic frame:
\begin{equation*}
{\zeta_{L_P}}^{-1}\,\zeta_{R}\,\zeta_{L_P}\; P \;\zeta_{L_P}^\dagger\,\zeta^\dagger_{R}\,{\zeta_{L_P}^\dagger}^{-1} =P \ .
\end{equation*}

But this road would not be convenient for us, since the generators of this group of transformations do not write in a nice form in spinor-helicity formalism. For instance the $J^0$ generator corresponding to the transformations~$\zeta_R$ reads
\begin{equation}
J^0=-\frac12\bigg(\lambda_1\frac{\partial}{\partial\lambda_1}-\lambda_2\frac{\partial}{\partial\lambda_2}-\tilde\lambda_1\frac{\partial}{\partial\tilde\lambda_1}+\tilde\lambda_2\frac{\partial}{\partial\tilde\lambda_2}+\mu_1\frac{\partial}{\partial\mu_1}-\mu_2\frac{\partial}{\partial\mu_2}-\tilde\mu_1\frac{\partial}{\partial\tilde\mu_1}+\tilde\mu_2\frac{\partial}{\partial\tilde\mu_2}\bigg) \ .
\end{equation}
One can see that it cannot be written in a compact form for $\lambda,\tilde{\lambda}$ and $\mu,\tilde{\mu}$, contrary to the operator~\eqref{Hop}. Let us instead consider another $\SU$ transformation, namely
\begin{equation}	\label{Utransf}
\bigg(\begin{array}{c}\lambda\\ \mu\end{array}\bigg)\to U\bigg(\begin{array}{c}\lambda\\ \mu\end{array}\bigg) \ ,\qquad
\Big(\,\tilde\lambda\;\;\tilde\mu\,\Big) \to \Big(\,\tilde\lambda\;\;\tilde\mu\,\Big)U^{\dagger} \ , \qquad\; \text{with }\ U\in \SU \ .
\end{equation}
Our massive momentum is always invariant under such transformations, independently of the frame:
\begin{equation}
\Big(\,\lambda\;\;\mu\,\Big)\,U^{\intercal} {U^{\intercal}}^{\dagger}\left(\!\begin{array}{c}\tilde{\lambda}\\\tilde{\mu}\end{array}\!\right) = \Big(\,\lambda\;\;\mu\,\Big)\left(\!\begin{array}{c}\tilde{\lambda}\\\tilde{\mu}\end{array}\!\right)= \lambda\tilde{\lambda}+\mu\tilde{\mu} \ .
\end{equation}

So these transformations $U$ are perfect candidates as massive LG transformations. Moreover, $U$ is actually a four-by-four matrix, which is an actual $\SU$ matrix composed with the two-by-two identity, \emph{i.e.}:
\begin{equation}	\label{Umatrix}
U=\Bigg(\begin{array}{cc}
	\alpha\, \mathbb{I}_2 	& \beta\, \mathbb{I}_2 \\
	-\beta^* \mathbb{I}_2\,	& \alpha^* \mathbb{I}_2 \\
\end{array}\Bigg)\ , \qquad \text{with }\ |\alpha|^2+|\beta|^2=1\ .
\end{equation}
So, in \eqref{Utransf}, $U$ is acting in the same way on both components $\lambda_1$ and $\lambda_2$ of the bi-spinor~$\lambda$, and in the same way on both components $\mu_1$ and $\mu_2$ of the bi-spinor~$\mu$. Then the infinitesimal generator $J_0$ for these transformations\footnote{
	A basis of generators of $\SU$ matrices is given by $J^0=\frac{\sigma^3}{2}=\begin{psmallmatrix}1&0\\0&-1\end{psmallmatrix}$, $J^+=\frac{\sigma^1+i\sigma^2}{2}=\begin{psmallmatrix}0&1\\0&0\end{psmallmatrix}$, $J^-=\frac{\sigma^1-i\sigma^2}{2}=\begin{psmallmatrix}0&0\\1&0\end{psmallmatrix}$, from which the respective differential operators are derived.
} reads
\begin{align*}
J^0 & = -\frac12\left(
	\lambda_1\bfrac{\partial\phantom{\lambda}}{\partial\lambda_1} +\lambda_2\bfrac{\partial\phantom{\lambda}}{\partial\lambda_2} -\tilde\lambda_1\bfrac{\partial\phantom{\lambda}}{\partial\tilde\lambda_1} -\tilde\lambda_2\bfrac{\partial\phantom{\lambda}}{\partial\tilde\lambda_2} -\mu_1\bfrac{\partial\phantom{\mu}}{\partial\mu_1} -\mu_2\bfrac{\partial\phantom{\mu}}{\partial\mu_2} +\tilde\mu_1\bfrac{\partial\phantom{\mu}}{\partial\tilde\mu_1} +\tilde\mu_2\bfrac{\partial\phantom{\mu}}{\partial\tilde\mu_2}\right) = \\ 
& = -\frac12\left(
	\lambda\bfrac{\partial\phantom{\lambda}}{\partial\lambda} -\tilde\lambda\bfrac{\partial\phantom{\lambda}}{\partial\tilde\lambda} -\mu\bfrac{\partial\phantom{\mu}}{\partial\mu} +\tilde\mu\bfrac{\partial\phantom{\mu}}{\partial\tilde\mu}\right) \ .
\end{align*}
So this operator recasts in a nice form in terms of $\lambda$ and $\mu$, analogous to that of the helicity operator~\eqref{Hop}. The same holds true for the other generators, and we can summarize:
\begin{align}
&
	J^0=-\frac12\left( \lambda\bfrac{\partial\phantom{\lambda}}{\partial\lambda} -\mu\bfrac{\partial\phantom{\mu}}{\partial\mu} -\tilde\lambda\bfrac{\partial\phantom{\lambda}}{\partial\tilde\lambda} +\tilde\mu\bfrac{\partial\phantom{\mu}}{\partial\tilde\mu}\right) \ , \label{J0U}\\
&
	J^+=-\mu\bfrac{\partial\phantom{\lambda}}{\partial\lambda} +\tilde{\lambda}\bfrac{\partial\phantom{\mu}}{\partial\tilde\mu} \ , \label{JpU}\\
&
	J^-=-\lambda\bfrac{\partial\phantom{\mu}}{\partial\mu} +\tilde{\mu}\bfrac{\partial\phantom{\lambda}}{\partial\tilde\lambda} \ . \label{JmU}
\end{align}

We underline that there is an isomorphic map between the $U$-transformations and the $\SU$ transformations~$\zeta_R$. It can be derived explicitly in the rest frame (and then extended through boosts to any frame), imposing the identity
\begin{equation*}
U\bigg(\begin{array}{c}\lambda\\ \mu\end{array}\bigg) = \bigg(\begin{array}{c}\zeta_R\lambda\\ \zeta_R\mu\end{array}\bigg)
\end{equation*}
component by component. So these $U$-transformations are full-fledged LG transformations, and we will legitimately use operators~(\ref{J0U}--\ref{JmU}) to constrain the amplitude. For a more formal derivation of such representation of the LG generators, the reader can check Section~2 and Appendix~A of~\cite{Conde:2016izb}.

\label{fakeLG}
We eventually remark that these groups of transformations, either $U$ or $\zeta_R$, are the physical LG transformations, but they are not the largest group of transformations that leave $P=\lambda\tilde{\lambda}+\mu\tilde{\mu}$ invariant. Indeed we can keep $\lambda,\,\tilde{\lambda}$ unchanged and scale $\mu,\,\tilde{\mu}$ as follows
\begin{equation} 	\label{scalemu}
\mu \longrightarrow t\,\mu \ , \qquad \tilde{\mu} \longrightarrow t^{-1}\mu \ ,
\end{equation}
or the other way round. Such transformations leave $P$ invariant, but they are not $\SU$ transformations, neither of the form~\eqref{Umatrix} nor of the form~\eqref{zetaR}. This ambiguity is related to the redundancy in our description of a time-like momentum in terms of null ones~\eqref{Paa}, as we have already stressed there: we can of course apply the respective massless LG transformations on each of the null momentum in the decomposition independently of the other; but this strictly depends on the given decomposition, while the actual massive LG transformation cannot depend on how we choose to decompose the massive momentum. So these additional transformations are not physical and we will have to demand the amplitude not to depend on them.

But let us first proceed to the analysis of the constraints given by the massive LG equations on the three-point amplitude.

\subsection{The massive three-point amplitude}\label{mass3ampl}

The starting point are the massive LG equations~\eqref{J0pmM}, which we rewrite here for the three-point amplitude:
\begin{align}
	J^0_jM_3\big(\{p_i\};\ldots, \sigma_j,\ldots\big) 	&= 	\sigma_j\,M_3\big(\{p_i\};\ldots, \sigma_j,\ldots\big) \ , \label{J0M3}\\
	J^\pm_jM_3\big(\{p_i\};\ldots, \sigma_j,\ldots\big) &= 	\sigma^\pm_j\,M_3\big(\{p_i\};\ldots, \sigma_j\pm 1,\ldots\big) \ . \label{JpmM3}
\end{align}
As we have remarked in advance around equations~\eqref{massrep}, the massive LG equations are such that the eq.~\eqref{J0M3} is an eigenvalue equation exactly as the helicity one~\eqref{J3M}, whereas the eq.s~\eqref{JpmM3} are relating different amplitudes. Of course, we would like to have a maximal number of differential equations for the same function, in order to hope to solve the system. 

The smart thing we can do is considering the amplitude where all massive particles are in the lowest value of their spin projection, $\sigma_i=-s_i$, so that the action of $J_-$ annihilates such amplitude:
\begin{equation}	\label{J-M3zero}
J_j^-M_3\big(\{p_i\};\ldots, -s_j,\ldots\big) = 0 \ .
\end{equation}
Then we have two simple equations for this amplitude for each massive particle: this last one, and the one corresponding to the action of~$J_0$~\eqref{J0M3}. With the operators~$J_i^+$ we can then raise the value of spin projections~$\sigma_i$, and obtain all the other amplitudes with higher values of~$\sigma_i$. Moreover, once we get to the highest value of the spin projection, $\sigma_i=+s_i$, we can act one more time with $J^+_i$ and again we annihilate the amplitude:
\begin{equation}	\label{JpM2s}
\big(J_j^+\big)^{2s_j+1}M_3\big(\{p_i\};\ldots, -s_j,\ldots\big) = 0 \ .
\end{equation}
This is yielding an additional constraint on this amplitude, but it is much more involved than~\eqref{J0M3} and~\eqref{J-M3zero}. So we will keep it for the end, considering for the moment only the simpler equations for $J_i^0$ and $J_i^-$.

Summarizing, we want to consider the three-point amplitudes involving massive particles that are in the lowest value of their spin component, $\sigma_i=-s_i$, as well as massless particles with arbitrary helicity, $h_i=\pm s_i$. Taking into account the expressions of the massive LG operators in spinor formalism~(\ref{J0U}-\ref{JmU}), together with the helicity equation~\eqref{HM3} for each massless legs, we have the following system of equations for this `lowest-component' amplitude:
\begin{equation}
\label{eqsM3ls}
\left\{
\begin{array}{ll}
H_j\,M^{\{a_i\}} = h_j\, M^{\{a_i\}}\ , 	&\quad
	\textrm{if }\ a_j=h_j\ ;				\vphantom{M^{\tilde{|}}}\\
&	\\
\begin{array}{l}
\!\!	J_j^0\,M^{\{a_i\}} = -s_{i}\, M^{\{a_i\}}\ ,	\vphantom{\frac{}{|}} \\
\!\!	J_i^-\,M^{\{a_i\}}=0\ ,							\vphantom{\big|}
\end{array}		&\quad
	\textrm{if }\ a_j=-s_j\ .
\end{array}	\right. 
\end{equation}
Again we have omitted the subscript of~$M$ to make notation lighter.

We now treat separately the cases with one, two, and three massive external states, from the simplest to the most involved. But before proceeding we discuss briefly the kinematic constraints coming from momentum conservation and on-shell conditions.

\subsubsection*{\refstepcounter{subsubsection}\thesubsubsection\hspace{2.3pt} Kinematic constraints: momentum conservation and on-shell conditions}

Since we use one pair of spinors for each massless momentum and two pairs for each massive one, we will need four pairs of spinors for the one-massive, two-massless case, five pairs for the two-massive, one-massless case, and six pairs for the three-massive case. Anyway, for all three cases momentum conservation turns into momentum conservation for (four, five, six) massless momenta. In general, momentum conservation for $m$ massive momenta and $n-m$ massless ones in our description is equivalent to momentum conservation of $n+m$ massless momenta.

We consider thus the general case of $n$ massless momenta, which means $n$ `non-tilded' and $n$ `tilded' bi-spinors. Out of $2n$ spinors we can build $\frac{1}{2}n\,(n-1)$ `angle' products~\eqref{ket} and $\frac{1}{2}n\,(n-1)$ square products~\eqref{bra}.

First, the massless on-shell condition for each of the $n$ momenta is automatically implemented thanks to spinor-helicity formalism, and this translates, as we have already seen in section~\ref{null3ampl}, into the geometrical statement that three bi-spinors cannot be linearly independent, \emph{i.e.}
\begin{equation}	\label{schouten}
\ket jk \lambda_i+\ket ki \lambda_j+\ket ij \lambda_k=0 \ .
\end{equation}
This fact goes under the name of Schouten identity. We can then choose for instance $\lambda_1$ and $\lambda_2$ as projecting directions, and use~\eqref{schouten} to express any of the angle products involving neither $\lambda_1$ nor $\lambda_2$ in terms of
\begin{equation}
\ket{1}{2}\, , \;	\ket{1}{i}\, , \; 	\ket{2}{i}\, , \qquad \text{with } i=3,\ldots,\,n \ .
\label{12i}
\end{equation}
These are $2\left(n\!-\!2\right)+1=2n\!-\!3$ independent variables.

We can then consider momentum conservation, which in spinor-helicity formalism reads
\begin{equation}
\label{1+n}
\sum_{i=1}^n\lambda_i\tilde\lambda_i=0 \ .
\end{equation}
If we contract this equation with $\lambda_1$ and $\lambda_2$ respectively, we obtain
\begin{equation}	\label{12til}
\tilde\lambda_1=-\sum_{i=3}^n \frac{\ket{i}{2}}{\ket12}\tilde\lambda_i \ ,	\qquad \text{and }\quad 
	\tilde\lambda_2=-\sum_{i=3}^n \frac{\ket{1}{i}}{\ket12}\tilde\lambda_i \ .	
\end{equation}
We see that in this way only the angle products that we have chosen as independent variables in~\eqref{12i} appear in the relations~\eqref{12til}. With them we can express all the square products involving either $\tilde\lambda_1$ or $\tilde\lambda_2$ in terms of
\begin{equation}
\bra{i}{j}	\qquad 	\text{with }\ i,j\neq 1,2 \ ,		\label{braij}
\end{equation}
which are $\frac12(n-2)(n-3)$ variables. When $n\!>\!5$, the variables~\eqref{braij} are not all independent since there are Schouten identities relating them, so we can further reduce the number of square products to $2(n\!-\!4)+\!1=2n\!-\!7$.

Thus, thanks to massless on-shell conditions and momentum conservation, we have reduced the total number of independent variables from the initial $n\left(n-1\right)$ to
\begin{equation}		\label{nindvar}
\left\{
\begin{array}{ccccll}
2n-3 	& + &
\frac12(n-2)(n-3) 	& = &
\frac12 n\,(n-1) & 
\qquad \text{if }\, n\leq5 		\vspace{9pt} \\
2n-3 	& + &
2n-7 	& = & 
2\left(2n-5\right) & 	
\qquad 	\text{if }\, n>5 		
\end{array}
\right. \ .	
\end{equation}

This conclusion is completely general, and holds for any kinematic process with $n$ conserved massless momenta. In our case, since we want to consider massive momenta, we have one additional condition per each massive particle, the massive on-shell condition~\eqref{Paa}.

If we refer to our counting of LG differential equations for massless and massive external particles~\eqref{eqsM3ls}, we can already compare the number of equation to the number of independent variables for each case. For three massless particles we had three equations and three independent variables. For one massive and two massless particles we have four equations from~\eqref{eqsM3ls}, and six independent variables from~\eqref{nindvar}, which further reduce to five because of one massive on-shell condition. For two massive and one massless particles, we have five equations and eight independent variables. And for three massive particles we have six equations and eleven variables.

Such an unfair comparison (clearly displayed in Table~\ref{numbvar}) could make us believe that we will hardly be able to completely determine the amplitude as in the fully massless case. Nonetheless, we are going to see how this is indeed possible.
\begin{table}
	\centering
	\renewcommand{\arraystretch}{1.4}
	\begin{tabular}{|R{3.7cm}|C{5mm}|C{5mm}|C{5mm}|C{5mm}|}
		\hline
		\# of massive legs\hspace*{1ex} 	& 0 	& 1 	& 2 	& 3 	\\
		\hline
		\# of LG equations\hspace*{1ex} 	& 3 	& 4 	& 5 	& 6 	\\
		\hline
		\# of indep. var.s\hspace*{1ex} 	& 3 	& 5 	& 8 	& 11	\\
		\hline
	\end{tabular}
	\caption{\label{numbvar}}
\end{table}

\subsubsection*{\refstepcounter{subsubsection}\thesubsubsection\hspace{2.3pt} One-massive two-massless amplitude}

We first consider the three-point amplitude with one massive particle and two massless one. We decide to parametrize the involved momenta through four pairs of spinors in the following way
\begin{equation}
\label{p1234}
P_1=\lambda_1\tilde\lambda_1+\lambda_4\tilde\lambda_4\ ,\qquad
p_2=\lambda_2\tilde\lambda_2\ ,\qquad
p_3=\lambda_3\tilde\lambda_3 \ ,
\end{equation}
with the mass condition reading
\begin{equation}
\ket14 \bra41 =  m^2 \ .	\label{masscond1}
\end{equation}

From the system~\eqref{eqsM3ls}, we have four equations for the three-point amplitude~$M^{h_{1},h_{2},-s_{3}}$, which in this section will be denote simply by~$M$. Using the expressions~(\ref{J0U}--\ref{JmU}) for the massive LG operators, and~\eqref{Hop} for the helicity operators, we can write 
\begin{equation}	\label{eqs1Mleg}
\begin{aligned}\!\!
\left(\lambda_1\bfrac{\partial\phantom{\lambda}}{\partial\lambda_1} 
	-\lambda_4\bfrac{\partial\phantom{\lambda}}{\partial\lambda_4} 
		-\tilde\lambda_1\bfrac{\partial\phantom{\lambda}}{\partial\tilde\lambda_1}  
			+\tilde\lambda_4\bfrac{\partial\phantom{\lambda}}{\partial\tilde\lambda_4}\right) M 
		&=+2s_{1}\, M \ ,	&&
	\left(\lambda_1\bfrac{\partial\phantom{\lambda}}{\partial\lambda_4} 
		-\tilde{\lambda}_4\bfrac{\partial\phantom{\lambda}}{\partial\tilde\lambda_1}\right) 
			M =0 \ ; \\
\left(\lambda_2\bfrac{\partial\phantom{\lambda}}{\partial\lambda_2} 
	-\tilde\lambda_2\bfrac{\partial\phantom{\lambda}}{\partial\tilde\lambda_2}\right) M 
		&=-2h_{2}\, M  \ ;	\vphantom{\Bigg|} \\
\left(\lambda_3\bfrac{\partial\phantom{\lambda}}{\partial\lambda_3} 
	-\tilde\lambda_3\bfrac{\partial\phantom{\lambda}}{\partial\tilde\lambda_3}\right) M
		&=-2h_{3}\, M \ .
\end{aligned}					
\end{equation}
As we have done in section~\ref{null3ampl}, since we know that the amplitude can only depend on $\SL$-invariant products of spinors, we can denote
\begin{equation} \label{12vars}
\!\!\!\!\begin{array}{clclclclclcl}
x_1=\ket23 	&\!\!\!,& x_2=\ket31 &\!\!\!,& x_3=\ket12 &\!\!\!,& x_4=\ket34 &\!\!\!,& x_5=\ket24 &\!\!\!,& x_6=\ket14 &\!\!\!, \\
y_1=\bra32 	&\!\!\!,& y_2=\bra13 &\!\!\!,& y_3=\bra21 &\!\!\!,& y_4=\bra43 &\!\!\!,& y_5=\bra42 &\!\!\!,& y_6=\bra41 &\!\!\!.
\end{array}
\end{equation}

Then we can use the chain rule to translate the differential operator in~\eqref{eqs1Mleg} in terms of these twelve variables, obtaining
\begin{equation}\label{PDE1m}
\begin{aligned}
	\left(x_3\partial_5-x_2\partial_6+y_6\tilde\partial_2-y_5\tilde\partial_3\right) M 
		&=0  \ , \\
	\left(x_2\partial_2+x_3\partial_3-x_5\partial_5-x_6\partial_6-y_2\tilde\partial_2-y_3\tilde\partial_3+y_5\tilde\partial_5+y_6\tilde\partial_6\right) M 
		&=2s_{1}\, M \ , \\
	\left(x_1\partial_1+x_3\partial_3+x_5\partial_5-y_1\tilde\partial_1-y_3\tilde\partial_3-y_5\tilde\partial_5\right) M 
	&=-2h_{2}\, M \ , \\
	\left(x_1\partial_1+x_2\partial_2+x_6\partial_6-y_1\tilde\partial_1-y_2\tilde\partial_2-y_6\tilde\partial_6\right) M 
	&=-2h_{3}\, M  \ ,
\end{aligned}
\end{equation}
where again we have used the same shorthand notations as in~\eqref{eqsm0M3} for partial derivatives with respect to $x_i$ and $y_i$. From the discussion of previous section, we know that only five variables over twelve are independent. It is convenient to choose $\lambda_2$ and $\lambda_3$ as reference directions in~\eqref{schouten} (and consequently $\tilde{\lambda}_2$ and $\tilde{\lambda}_3$ in~\eqref{12til}), expressing in this way all the variables in terms of $x_1$, $x_2$, $x_3$, $x_4$, $x_5$ (this choice is completely arbitrary, but it will reveal as the most convenient to express the result in its simplest form). We get the following expressions for the other variables:
\begin{equation}	\label{1mcons}
x_6=-\frac{x_1x_4+x_2x_5}{x_3}\ ;	\qquad
\begin{array}{lll}
\displaystyle 	y_1=\frac{m^2}{x_1}\ , &\,\
\displaystyle 		y_2=\frac{m^2x_5}{x_1x_4}\ , &\,\
\displaystyle 			y_3=-\frac{m^2}{x_1x_4}\frac{x_1x_4+x_2x_5}{x_3}\ , \\[2ex]
\displaystyle 	y_4=\frac{m^2}{x_4}\ , &\,\
\displaystyle 		y_5=\frac{m^2x_2}{x_1x_4}\ , &\,\
\displaystyle 			y_6=\frac{m^2x_3}{x_1x_4} \ .
\end{array}
\end{equation}

Now we can use the chain rule the other way round to express the system in terms only of the chosen independent variables, obtaining
\begin{equation}	\label{TDE1m}
\begin{aligned}
\partial_5 M &=0\ , \\
\big(x_2\partial_2+x_3\partial_3-x_5\partial_5\big)\,M 	& = +2s_{1}M\ , \\
\big(x_1\partial_1+x_3\partial_3+x_5\partial_5\big)\,M		&=-2h_{2} M\ , \\
\big(x_1\partial_1+x_2\partial_2\big)\,M 	& = -2h_{3} M\ .
\end{aligned}
\end{equation}
Indeed the chain rule for the changes of variable~\eqref{1mcons} yields
\begin{equation}
\begin{aligned}
& x_1\partial_1 \longrightarrow
	x_1\partial_1 -\frac{x_1x_4}{x_3}\,\partial_6 -\frac{m^2}{x_1}\,\tilde{\partial}_1 -\frac{m^2x_5}{x_1x_4}\,\tilde{\partial}_2 +\frac{m^2x_2x_5}{x_1x_4x_3}\,\tilde{\partial}_3 -\frac{m^2x_2}{x_1x_4}\,\tilde{\partial}_5 -\frac{m^2x_3}{x_1x_4}\,\tilde{\partial}_6 \ ,\\
& x_2\partial_2 \longrightarrow
	x_2\partial_2 -\frac{x_2x_5}{x_3}\,\partial_6 -\frac{m^2x_2x_5}{x_1x_4x_3}\,\tilde{\partial}_3 +\frac{m^2x_2}{x_1x_4}\,\tilde{\partial}_5 \ ,\\
& x_3\partial_3 \longrightarrow
	x_3\partial_3 +\frac{x_1x_4+x_2x_5}{x_3}\,\partial_6 +\frac{m^2}{x_1x_4}\frac{x_1x_4+x_2x_5}{x_3}\,\tilde{\partial}_3 +\frac{m^2x_3}{x_1x_4}\,\tilde{\partial}_6 \ ,\\
& x_4\partial_4 \longrightarrow
	x_4\partial_4 -\frac{x_1x_4}{x_3}\,\partial_6 -\frac{m^2x_5}{x_1x_4}\,\tilde{\partial}_2 +\frac{m^2x_2x_5}{x_1x_4x_3}\,\tilde{\partial}_3 -\frac{m^2}{x_4} -\frac{m^2x_2}{x_1x_4}\,\tilde{\partial}_5 -\frac{m^2x_3}{x_1x_4}\,\tilde{\partial}_6 \ ,\\
& x_5\partial_5 \longrightarrow
	x_5\partial_5 -\frac{x_2x_5}{x_3}\,\partial_6 +\frac{m^2x_5}{x_1x_4}\,\tilde{\partial}_2 -\frac{m^2x_2x_5}{x_1x_4x_3}\,\tilde{\partial}_3 \ .
\end{aligned}
\end{equation}
If we substitute these rules into~\eqref{TDE1m}, we get precisely the system~\eqref{PDE1m}. Notice that it is not granted at all, that the operators in~\eqref{PDE1m} can be obtained from the operators~\eqref{TDE1m}, namely containing only differentials of the independent variables, upon the application of the constraints~\eqref{1mcons}. Some magic is happening, reflecting the compatibility of the constraints~\eqref{1mcons}, coming from Poincar\'e invariance, with our LG differential operators.\footnote{
	If we take for instance the differential operator $x\partial_x+y\partial_y$ with the constraint $y=x^2$, it cannot be expressed as $x\partial_x$, which instead with this constraint is giving $x\partial_x\rightarrow x\partial_x+2y\partial_y$. It would be interesting to explicitly show why the constraints coming from Poincar\'e invariance happen to be compatible with the LG differential operators, at least in all the cases discussed here. 
} It can be taken as a confirmation of the consistency of our treatment.

Let us now solve the system~\eqref{TDE1m}. The first equation tells us that the amplitude does not depend on $x_5$, so that the other three equations yield exactly the same system as in the massless case~\eqref{eqsm0M3}! (with $h_{1}$ replaced by $-s_{1}$) Moreover, we note that $x_4$ is not appearing in the equations, so that there is no constraint at all on the dependency of the amplitude on that variable. Thus, the most general solution for the one-massive-leg lowest-component three-point amplitude is
\begin{equation}	 	\label{sol1m}
\begin{aligned}
M^{-s_1,\,h_2,h_3} = \; &
x_1^{-s_{1}-h_{2}-h_{3}}\, x_2^{h_{2}-h_{3}+s_{1}}\, x_3^{h_{3}-h_{2}+s_{1}} \: f_1(x_4) \\
= \; &	{\ket12}^{h_{3}-h_{2}+s_{1}}\: {\ket23}^{-s_{1}-h_{2}-h_{3}}\: {\ket31}^{h_{2}-h_{3}+s_{1}}\; f_1\big(\ket14\big) \ ,
\end{aligned}
\end{equation}
where $f_1$ is an arbitrary function, which depends on $\ket14$ and on other parameters of the interaction, like the mass~$m$ and the coupling constant~$g$. The mass dimension of $f_1$ is fixed since the three-point amplitude must have mass dimension equal to one. Then we can factorize the dimensionful part of $f_1$,
\begin{equation}	\label{tf1}
f_1(\ket14)=g\,m^{1-[g]-s_1+h_2+h_3}\,\tilde{f}_1\Big(\textstyle\bfrac{\ket14}{m}\Big) \ ,
\end{equation}
where $g$ is the coupling constant of the interaction and $[g]$ is its mass dimension. In this way the function $\tilde{f}_1$ is now dimensionless, depending only on the dimensionless argument~$\ket14/m$. Furthermore, we will argue that $\tilde{f}_1$ is just a constant.

Indeed, we notice that the argument $\ket14$ is related to our ambivalent choice of decomposition of the massive momentum~$P_1$. Following the considerations on page~\pageref{fakeLG}, we can apply the transformation~\eqref{scalemu} on $\lambda_4$, and make it scale, whereas we leave $\lambda_{1}$ untouched: in this way, requiring the amplitude to be independent of such unphysical scaling is equivalent to demanding $\tilde{f}_1$ in~\eqref{tf1} to be constant. We can then absorb it into the coupling constant, obtaining the following final expression for the physical one-massive leg three-point amplitude~\cite{Conde:2016vxs}:
\begin{equation}	\label{M1madim}
M^{-s_1,\,h_2,h_3} = g\,m^{1-[g]-s_1+h_2+h_3}\,{\ket12}^{h_{3}-h_{2}+s_{1}}\: {\ket23}^{-s_{1}-h_{2}-h_{3}}\: {\ket31}^{h_{2}-h_{3}+s_{1}}\ ,
\end{equation}
which is thus completely determined by Poincar\'e invariance, exactly as its massless sibling. But contrarily to its massless counterpart, this amplitude is non-zero even for real kinematics, representing the decay of a massive particle into two massless ones. So, it constitutes a full non-perturbative result, being derived only from symmetry-based considerations.

The attentive reader might remember now of the cumbersome constraint~\eqref{JpM2s}, and wonder how it could further constrain the amplitude, as it is already `fully' determined. Actually, if one acts $2s_1+1$~times on the amplitude~\eqref{M1madim} with the spin-raising operator for particle~1, \emph{i.e.}
\begin{equation}
J_1^+ = -\lambda_4\bfrac{\partial\phantom{\lambda}}{\partial\lambda_1} +\tilde{\lambda}_1\bfrac{\partial\phantom{\lambda}}{\partial\tilde\lambda_4} \ ,
\end{equation}
and require the result to vanish, the following condition on the helicities of the two massless particles is obtained~\cite{Conde:2016vxs}:
\begin{equation}	\label{h2h3cond}
h_2-h_3 = \{-s_1,\, -s_1+1,\, \ldots,\, s_1-1,\, s_1\} \ .
\end{equation}
Such condition on the difference of the helicities of the two massless particles is a very basic relation descending from the conservation of angular momentum. Indeed, we can quickly see the case where the difference between the helicities is maximized, that is the frame where the spatial momenta of the three particles are aligned:
\begin{equation*}
\bfrac{\vec{P}_1}{|\vec{P}_1|}=\pm\bfrac{\vec{p}_2}{|\vec{p}_2|}=\mp\bfrac{\vec{p}_3}{|\vec{p}_3|} \ .
\end{equation*}
Then, from angular momentum conservation, $\vec{J}_1+\vec{J}_2+\vec{J}_3=0$, and using the definition of helicity~\eqref{defHelicity}, we have
\begin{equation*}
\mp\bfrac{\vec{J}_3\cdot\vec{p}_3}{|\vec{p}_3|} = (-\vec{J}_1-\vec{J}_2)\cdot\bfrac{\vec{P}_1}{|\vec{P}_1|}\ \Rightarrow\ 
	\mp h_3 = s_1 \mp \bfrac{\vec{J}_2\cdot\vec{p}_2}{|\vec{p}_2|} = s_1 \mp h_2\ \Leftrightarrow\ 
		|h_2-h_3|=s_1 \ ,
\end{equation*}
in agreement with~\eqref{h2h3cond}.

We conclude this section with a nice, straightforward application of the formula~\eqref{sol1m}. A renowned fact in particle physics is the impossibility of a massive vector boson to decay into two photons, which goes under the name of Landau-Yang theorem~\cite{Landau:1948kw,Yang:1950rg}. We can indeed take the result~\eqref{sol1m}, set $s_1=1$, and consider the two cases where the helicities of the massless spin-1 particles have either the same sign or opposite sign:
\begin{equation*}
M^{-1,\,\pm1\,\pm1} = f_1\; \bfrac{\ket12\ket31}{\ket23} \big(\ket23\big)^{\mp 2}\ ; \qquad
	M^{-1,\,\pm1\,\mp1} = f_1\; \bfrac{\ket12\ket31}{\ket23}\bigg(\bfrac{\ket12}{\ket31}\bigg)^{\mp 2} \ .
\end{equation*}
You can see that in both case, if we switch particle 2 and particle 3, the amplitude flips sign. But since particles 2 and 3 are identical bosonic particles, their exchange should not affect the amplitude. We conclude that this amplitude has to be zero. The simplicity and shortness of this proof should be appreciated, compared to traditional derivations of the Landau-Yang theorem. Moreover, notice that with the formula~\eqref{sol1m} the statement can be easily generalized to massless particles of higher spin: a spin-1 massive particle cannot decay into two massless identical bosonic particles (\emph{i.e.}: of any arbitrary integer spin).

\subsubsection*{\refstepcounter{subsubsection}\thesubsubsection\hspace{2.3pt} Two-massive one-massless amplitude}

The case of two massive and one massless particles goes in a completely analogous way as the one we have just considered. We take
\begin{equation}
\label{p12345}
P_1=\lambda_1\tilde\lambda_1+\lambda_4\tilde\lambda_4\ ,\qquad
P_2=\lambda_2\tilde\lambda_2+\lambda_5\tilde\lambda_5\ ,\qquad
p_3=\lambda_3\tilde\lambda_3 \ ,
\end{equation}
with the mass conditions
\begin{equation}
\ket14 \bra41 =  {m_1}^2\ , \quad \ket25 \bra52 =  {m_2}^2 \ .	\label{masscond2}
\end{equation}

For the amplitude where the two massive particles are in their lowest spin component, the system of LG equations reads
\begin{align}
\left(\lambda_1\bfrac{\partial\phantom{\lambda}}{\partial\lambda_1} -\lambda_4\bfrac{\partial\phantom{\lambda}}{\partial\lambda_4}
	-\tilde\lambda_1\bfrac{\partial\phantom{\lambda}}{\partial\tilde\lambda_1}  +\tilde\lambda_4\bfrac{\partial\phantom{\lambda}}{\partial\tilde\lambda_4}\right)\! M &=+2s_{1}\, M\ , \quad\
		\left(\lambda_1\bfrac{\partial\phantom{\lambda}}{\partial\lambda_4} -\tilde{\lambda}_4\bfrac{\partial\phantom{\lambda}}{\partial\tilde\lambda_1}\right)\! M =0 \  ;\ \nn
\left(\lambda_2\bfrac{\partial\phantom{\lambda}}{\partial\lambda_2} -\lambda_5\bfrac{\partial\phantom{\lambda}}{\partial\lambda_5}
	-\tilde\lambda_2\bfrac{\partial\phantom{\lambda}}{\partial\tilde\lambda_2}  +\tilde\lambda_5\bfrac{\partial\phantom{\lambda}}{\partial\tilde\lambda_5}\right)\! M &=+2s_2\, M\ , \quad\
		\left(\lambda_2\bfrac{\partial\phantom{\lambda}}{\partial\lambda_5} -\tilde{\lambda}_5\bfrac{\partial\phantom{\lambda}}{\partial\tilde\lambda_2}\right)\! M =0 \  ;\ \nn
\left(\lambda_3\bfrac{\partial\phantom{\lambda}}{\partial\lambda_3} -\tilde\lambda_3\bfrac{\partial\phantom{\lambda}}{\partial\tilde\lambda_3}\right)\! M &=-2h_{3}\, M \ .
\label{eqs2Mleg}
\end{align}

So we have five equations, and from the five pairs of spinors in~\eqref{p12345} we can form twenty spinor products, which reduce to eight independent ones after applying the kinematic constraints~\eqref{nindvar}. We choose as independent variables the following ones,
\begin{equation}
\begin{array}{llll}
	x_1=\ket23 \ , \quad & x_2=\ket31  \ , \quad & x_3=\ket12  \ , \quad & x_4=\ket14  \ , \vphantom{\frac{}{\big|}}\\
	x_5=\ket25 \ , \quad & x_6=\ket24  \ , \quad & x_7=\ket15  \ , \quad & y_8=\,\bra54 \ .
\end{array}
\end{equation}
so seven angle-products and one square-product. Again, the LG equations can be perfectly recast in terms of differential of the independent variables only, obtaining
\begin{equation}
\begin{aligned}
x_1\partial_1 M &= \left(s_{2}-s_{1}-h_{3}\right) M \ , \phantom{\Big|} \quad & 
\partial_6 M &= 0 \ , \\
x_2\partial_2 M &= \left(s_{1}-s_{2}-h_{3}\right) M \ , \phantom{\Big|} \quad & 
\partial_7 M &= 0 \ , \\
\big(x_3\partial_3+y_8\tilde{\partial}_8\big)M & =\left(s_{1}+s_{2}+h_{3}\right)M \ .
\end{aligned}
\end{equation}
The most general solution to this system is~\cite{Conde:2016vxs}
\begin{equation} 	\label{sol2m}
\begin{aligned}
M^{-s_{1},-s_{2},\,h_{3}} &= 
	x_1^{s_{2}-s_{1}-h_{3}}\: x_2^{s_{1}-s_{2}-h_{3}}\: x_3^{s_{1}+s_{2}+h_{3}}\; f_2\Big(x_4,\,x_5,\,\frac{y_8}{x_3}\Big) \\
&= 
	{\ket12}^{s_{1}+s_{2}+h_{3}}\: {\ket31}^{s_{1}-s_{2}-h_{3}}\: {\ket23}^{s_{2}-s_{1}-h_{3}}\; f_2\bigg(\!{\ket14},{\ket25},\bfrac{\bra54}{\ket12}\bigg)	\ .
\end{aligned}
\end{equation}
We recognize again a factor carrying the proper Lorentz-wise scaling, again similar to that of the massless amplitude, together with an undetermined function $f_2$ of three variables. We can repeat the dimension-based considerations of~\eqref{tf1}, and rewrite $f_2$ as
\begin{equation}
\label{tf2}
f_2\bigg(\!{\ket14},{\ket25},\frac{\bra54}{\ket12}\bigg)= g\,m_{1}^{1-[g]-s_{1}-s_{2}+h_{3}}
\tilde{f}_2\bigg(\frac{\ket14}{m_{1}},\frac{\ket25}{m_{2}},\frac{\bra54}{\ket12}\,;\frac{m_{2}}{m_{1}}\bigg) \ ,
\end{equation}
so that $\tilde{f}_2$ is dimensionless. We notice that the first arguments of $f_2$ (or $\tilde{f}_2$) are the angle-products related to the mass, so we can expect the function not to depend on them, because of the argument about the non-physical scaling of $\lambda_4,\tilde{\lambda}_4$ (and $\lambda_5,\tilde{\lambda}_5$), that we have already applied to $f_1$; but the third argument contains actual kinematic information. 

We can then resort to the constraints of the form~\eqref{JpM2s}. We have two of them in the present case, \emph{i.e.}
\begin{equation}	\label{J+12}
\big(J^+_1\big)^{2s_{1}+1}M^{-s_{1},-s_{2},\,h_{3}}=0 \ ,\qquad
	\big(J^+_2\big)^{2s_2+1}M^{-s_{1},-s_{2},\,h_{3}}=0 \ ,
\end{equation}
whose respective actions on the amplitude~\eqref{sol2m}\footnote{
	See Appendix B of the original paper~\cite{Conde:2016vxs} for details of the calculation.
} determine the following two rational expressions for $f_2$:
\begin{align}
	\label{f2J+1}
f_2&=\sum_{k=0}^{2s_{1}} c^{(1)}_k\!\big(\ket14,\ket25\big)\: \bigg(1+\frac{\ket14\ket25}{{m_2}^{\!2}\vphantom{\big|}}\bfrac{\bra54}{\ket12}\bigg)^{\!s_{1}+s_{2}+h_{3}-k} \ , \\
	\label{f2J+2}
f_2&=\sum_{k=0}^{2s_{2}} c^{(2)}_k\!\big(\ket14,\ket25\big)\: \bigg(1+\frac{\ket14\ket25}{{m_1}^{\!2}\vphantom{\big|}}\bfrac{\bra54}{\ket12}\bigg)^{\!s_{1}+s_{2}+h_{3}-k}\ .
\end{align}
The coefficients $c_k^{(1)},\;c_k^{(2)}$ are still undetermined functions of their arguments, but now they depend only on the `mass' variables, and so, requiring the amplitude to be invariant under the non-physical scaling of $\lambda_4,\tilde{\lambda}_4$ ($\lambda_5,\tilde{\lambda}_5$) independently of $\lambda_1,\tilde{\lambda}_1$ ($\lambda_2,\tilde{\lambda}_2$), we obtain that they must be constants. Notice that on the other hand the remaining combinations of variables appearing as the power bases in~(\ref{f2J+1}-\ref{f2J+2}) are invariant under the non-physical scalings of $\lambda_4,\tilde{\lambda}_4$ and $\lambda_5,\tilde{\lambda}_5$, consistently.

We can go further realizing that these two expressions for $f_2$ have to be equivalent, since they describe the same function. Let us first assume different masses, $m_1\neq m_2$. Then, if we develop both expressions~(\ref{f2J+1}-\ref{f2J+2}) in powers of the common variable~$\bfrac{\bra54}{\ket12}$, and require the coefficients of equal powers to match, we will be forced to set to zero some of the constants~$c_k^{(i)}$. And in particular, when $|h_3|>s_1+s_2$, all of them will have to be zero. Again we find a constraint on the possible values of the helicity of the massless particle, namely~\cite{Conde:2016vxs}
\begin{equation}
h_3=\{-s_1-s_2,\,-s_1-s_2+1,\,\ldots,\,s_1+s_2-1,\,s_1+s_2\}\ ,
\end{equation}
and again for a case where the amplitude is physical, \emph{i.e.} non-zero for real momenta, as the masses are different.

Let us then consider the case where the masses are equal, which corresponds to a kinematically forbidden process. Up to truncating the longer of the two series, the expressions~\eqref{f2J+1} and~\eqref{f2J+2} are automatically matching, without need of any restriction on $h_3$. This parallels the massless case, where the amplitude is also forbidden for real kinematics, and indeed we had no constraints on the values of the helicities.

So, we have completely determined also the three-point amplitude with one massless and two massive legs, up to \emph{now several} constants, corresponding to some different kinds of coupling. But how many of them? If we take for instance $s_1\leq{s_2}$, we can convince ourselves that from the matching of~\eqref{f2J+1} and~\eqref{f2J+2} the following number of surviving (non-zero) constants~$c_k$ is given, depending on the values of the helicity of the massless particle~\cite{Conde:2016vxs}:
\begin{equation}
\# \ \text{of couplings}\ = \left\{
\begin{array}{llrl}
	s_{1}+s_{2}-h_{3}+1	&\qquad\textrm{if }\ & {s_2-s_1}&\leq{h_3}\leq{s_2+s_1} 	\\
		2s_{1}+1  \vphantom{\Big|} &\qquad\textrm{if }\ & {-s_2+s_1}&\leq{h_3}\leq{s_2-s_1} \vphantom{\Big|} 	\\
			s_{1}+s_{2}+h_{3}+1	&\qquad\textrm{if }\ & {-s_2-s_1}&\leq{h_3}\leq{-s_2+s_1} 		\\
\end{array}\right. \ ,
\end{equation}
which is always no more than~$2s_1+1$, with~$s_1\equiv\mathrm{min}\{s_1,s_2\}$.

Let us conclude with an example, to see in practice how these different couplings can arise. Consider the QED three-point vertex, that is two massive spin-$\tfrac{1}{2}$ fermions interacting with a massless vector boson. Here the electrons/positrons have the same mass and same spin, so we are not facing a physical amplitude representing the decay of a massive particle, but nevertheless we have a three-point vertex which intervenes in intermediate steps of the perturbative calculation of physical higher points amplitudes.

Let us take both fermions in their lowest spin component~$-\tfrac{1}{2}$, and consider a photon of helicity~$-1$. The formul\ae~\eqref{sol2m} and~\eqref{f2J+1}, with~$s_1=s_2=\tfrac12$ and~$h_3=-1$, yield
\begin{equation}	\label{112vertex}
M^{-\frac12,-\frac12,-1} = e\,m^{-1-[e]}\;{\ket23}{\ket31}\; \frac{c_0+c_1+c_0\,\xi}{1+\xi} \ , \qquad \text{with }\ \xi=\frac{\ket14\ket25}{m^{2\vphantom{\frac{a}{}}}}\bfrac{\bra54}{\ket12}\ ,
\end{equation}
and where we have renamed the coupling constant~$e$, foreseeing future identification with the electromagnetic coupling. We see that we have indeed two independent constants and two different functional structures. 

If we want to construct the same amplitude through a Lagrangian approach, we realize that in the Lagrangian of QED we have just one three-point vertex, \emph{i.e.}~$e\,\bar{\psi}\gamma^{\mu} A_{\mu}\psi$, with a single coupling constant related to the electric charge of the electron. However, the quantum corrected three-point vertex exhibits, already at one loop, a second piece, proportional to $\gamma^{\mu\nu}=\frac{i}{2}[\gamma^\mu,\gamma^\nu]$:
\begin{equation*}
\Gamma^\mu_{loops}= \gamma^\mu\,G_1(p_3^2) +\frac{i}{2m}\gamma^{\mu\nu}{p_3}_\nu\:G_2(p_3^2)\ .
\end{equation*}
This latter quantum-generated term is responsible for the anomalous gyromagnetic moment of the electron\footnote{
	This is standard material of any textbook on Quantum Field Theory. Check for instance Chapter 6 of Peskin-Schroeder's book~\cite{Peskin1995qft}, or Sections 10.6 and 11.3 of Weinberg's book~\cite{Weinberg:1995mt}.
}. 

Considering the following expressions for the polarization vector of the photon of negative helicity\footnote{
	The polarization vector is defined in spinor helicity formalism by means of an arbitrary reference spinor, that here we choose to be $\lambda_4$. This ambiguity of definition is related to gauge transformations, and one particular choice corresponds to a gauge fixing. The final answer is gauge invariant, and so not depending on this choice. For instance, it can be easily check that choosing $\tilde{\lambda}_5$ instead of $\tilde{\lambda}_4$ as reference spinor would not change the result.
} and for the wave functions of the Dirac fermions in spinor/helicity formalism,
\begin{equation}
\sigma^\mu\epsilon^-_\mu(p_3)=\frac{\lambda_3\,\tilde{\lambda}_4}{\bra{3}{4}} \ , \quad
	\bar{v}_-(P_1)=\left({\tfrac{\ket14}{m}}\,\tilde\lambda_4,\; \lambda_1\right) \ ,\quad
		u_-(P_2)=\left(\!\!\begin{array}{c}{\frac{\ket25}{m}}\,\tilde\lambda_5 \\ \lambda_2 \end{array}\!\!\right) \ ,
\end{equation}
we can write the electron-electron-photon three point amplitude from the QED renormalized cubic Lagrangian, ${\cal L}_{QED}^{ren}=\bar{\psi}\big(e\,\gamma^{\mu} A_{\mu}+\frac{eg}{2m}\,i\gamma^{\mu\nu}F_{\mu\nu}\big)\psi$, which is taking into account the quantum contributions. We obtain
\begin{equation}	\label{112vertexQED}
\begin{aligned}
M^{-\frac12,-\frac12,-1}&=\bar{v}_-(P_1)\left(e\,\gamma^{\mu}+\frac{eg}{2m}\,i\gamma^{\mu\nu}{p_3}_\nu\right)\epsilon^-_{\mu}(p_3)\,u_-(P_2) \\
&=\frac{e}{m}\;\ket23\ket31\left(\frac{-\xi}{1+\xi}+\frac{g}{2}\right) \ ,
\end{aligned}
\end{equation}
which is precisely matching the expression~\eqref{112vertex}, once we recall that the electromagnetic coupling is dimensionless. To complete the matching we have to choose $c_0=\frac{g}{2}-1$ and $c_1=1$.

We hope with this example to have shown how different Lorentz structures can arise for the same external particle content, and this independently of the adopted formalism.

\subsubsection*{\refstepcounter{subsubsection}\thesubsubsection\hspace{2.3pt} Three-massive amplitude}

The three massive case is completely analogous to the previously considered cases, just more involved due to the increasing number of variables, so we review it very briefly. We have here three pairs of spinors, which describe the three massive momenta,
\begin{equation*}	\label{p123456}
\begin{array}{rclclc}
&	P_1=\lambda_1\tilde\lambda_1+\lambda_4\tilde\lambda_4\ ,  &\hspace*{1em}&
		P_2=\lambda_2\tilde\lambda_2+\lambda_5\tilde\lambda_5\ , &\hspace*{1em}&
			P_3=\lambda_3\tilde\lambda_3+\lambda_6\tilde\lambda_6\ , \vphantom{\frac{}{\big|}}\\
\text{with }\hspace*{1em} &
	\ket14\bra41={m_{1}}^2\ , &&
		\ket25\bra52={m_{2}}^2\ , &&
			\ket36\bra36={m_{3}}^2\ ;
\end{array}	
\end{equation*}
thus we have thirty spinor products, eleven of which are independent. Since we have six LG equations in the system~\eqref{eqsM3ls} for three massive legs, then we expect the amplitude to depend on an undetermined function of five arguments. From the previous examples we can already guess that three of these arguments will be the angle-products related to the mass (that we are able to get rid of by imposing the non-dependence of the amplitude on the non-physical scaling), whereas the other two arguments would contain some square-products. Indeed, the final result is~\cite{Conde:2016vxs}:
\begin{align}	\label{sol3m}
& M^{-s_{1},-s_{2},-s_{3}} = \\
&\qquad = 
	{\ket12}^{s_{1}+s_{2}-s_{3}}\,{\ket31}^{s_{3}+s_{1}-s_{2}}\, {\ket23}^{s_{2}+s_{3}-s_{1}}\; f_3\bigg(\!{\ket14},{\ket25},{\ket36},\bfrac{{\bra54}}{{\ket12}},\bfrac{{\bra46}}{{\ket31}}\bigg) \ .	\nonumber
\end{align}
We find again the by now usual pre-factor embodying the LG scaling, and then the expected undetermined function of the remaining five variables, which we can again make dimensionless as in the previous cases: 
\begin{equation}	\label{tf3}
f_3=g\:m_{1}^{1-[g]-s_{1}-s_{2}-s_{3}}
\tilde{f}_3\bigg(\frac{\ket14}{m_{1}},\frac{\ket25}{m_{2}},\frac{\ket36}{m_{3}},\bfrac{{\bra54}}{{\ket12}},\bfrac{{\bra46}}{{\ket31}}\,;\frac{m_{2}}{m_{1}},\frac{m_{3}}{m_{1}}\bigg) \ .
\end{equation}

We can now expect this function to be specified by applying the spin-raising operators as for the two-massive case~\eqref{J+12}, which is true, even if more complicated. Indeed the three spin-raising operators, corresponding to particle 1, 2 and 3, are in this case
\begin{equation*}
J^+_1 = -\lambda_4\bfrac{\partial\phantom{\lambda}}{\partial\lambda_1} +\tilde{\lambda}_1\bfrac{\partial\phantom{\lambda}}{\partial\tilde\lambda_4}\ , \qquad
	J^+_2 = -\lambda_5\bfrac{\partial\phantom{\lambda}}{\partial\lambda_2} +\tilde{\lambda}_2\bfrac{\partial\phantom{\lambda}}{\partial\tilde\lambda_5}\ , \qquad
		J^+_1 = -\lambda_6\bfrac{\partial\phantom{\lambda}}{\partial\lambda_3} +\tilde{\lambda}_3\bfrac{\partial\phantom{\lambda}}{\partial\tilde\lambda_6}\ .
\end{equation*}
As you can see, the second operator acts only on the argument containing~$\tilde{\lambda}_5$ in the function~\eqref{tf3}, the third operator acts only on the argument containing~$\tilde{\lambda}_6$, whereas the first operator acts on both of them. This makes the solution to the equation~$(J^+_i)^{2s_{i}+1}M=0$ easy to find for $i=2,3$, but vary hard for $i=1$.

For $i=2,3$ respectively, the following expressions can be obtained~\cite{Conde:2016vxs}:
\begin{align}
f_3\Big(\ldots;\,{\textstyle\bfrac{\bra54}{\ket12},\bfrac{\bra46}{\ket31}}\Big) &=
	\sum_{k=0}^{2s_2} c^{(2)}_k\Big(\ldots;\,{\textstyle\bfrac{\bra46}{\ket31}}\Big)\; \left({\textstyle\ket25\bfrac{\bra54}{\ket12}+\ket36\bfrac{\bra46}{\ket31}+\bfrac{{m_1}^2}{\ket14}}\right)^{s_1-s_2-s_3+k} 	\nn
&=
	\sum_{k=0}^{2s_3} c^{(3)}_k\Big(\ldots;\,{\textstyle\bfrac{\bra54}{\ket12}}\Big)\; \left({\textstyle\ket25\bfrac{\bra54}{\ket12}+\ket36\bfrac{\bra46}{\ket31}+\bfrac{{m_1}^2}{\ket14}}\right)^{s_1-s_2-s_3+k}\ , \nonumber
\end{align}
where the dots stand for the variables~$\ket14,\ket25,\ket36$, which we know that the amplitude will be eventually independent of. But you see that in this case the coefficients~$c^{(2)}_k$ and~$c^{(3)}_k$ are not necessarily constants, since they depend also on a square-product, $\bfrac{\bra46}{\ket31}$ and~$\bfrac{\bra54}{\ket12}$ respectively. Of course, if we require the matching of these two different expressions of the same function, and impose the additional constraint coming from the action of $J^+_1$, we could in principle fully specify the form of~$f_3$, and extract as well some restrictions on the allowed spins. But in practice this is unfortunately too cumbersome, and it is unlikely possible to obtain a final expression for arbitrary spins, as for $f_2$~(\ref{f2J+1}-\ref{f2J+2}). However, it is (easily) feasible to work it out case by case, with given (little) values of the spins.

We conclude this section remarking that also this amplitude will eventually depend on several constants. If we take $s_1$ to be the highest of the three spins, then we can be quickly convinced that the number of constants cannot exceed $s_2\cdot{s_3}$. Of course, it would be interesting to precisely determine this number.

\section[BCFW recursion relations]{Britto-Cachazo-Feng-Witten recursion relations}	\label{BCFW}

Recursion relations for scattering amplitudes are in general relations connecting $n$-point amplitudes to lower-point ones, that can be thus applied recursively to construct arbitrarily-high-point amplitudes from lower-point information. You see that if we dispose of such a powerful tool, then the three-point amplitudes, which we have determined in chapter~\ref{3pAmpli}, constitute the fundamental starting point from which we would be able to recursively construct any other higher-point amplitude.

Before proceeding, we clarify that recursion relations are (so far) based on arguments that are valid order by order in the perturbative expansion. So, the methods we will discuss in this chapter are not non-perturbative, as those discussed in the previous one, which were based on symmetries.

The first principles that we will use to derive the on-shell recursion relations are locality, analyticity, unitarity.
\begin{description}
	
	\item[Locality] enters the discussion through the cluster decomposition principle, which we have introduced in Chapter~\ref{preliminaries}, page~\pageref{cluster}. This assures that the amplitude, once we have singled out the delta function of momentum conservation, exhibits no other delta-like singularities. Yet, locality of interaction will not be explicitly manifest in this context, as it is in a Lagrangian formulation of quantum field theory; actually, it can be violated in some intermediate steps\footnote{
		In the Lagrangian approach, the price for manifest locality is the gauge redundancy: Feynman diagrams are gauge dependent, even if the final results, after precise cancellations among different terms, is of course gauge invariant. In BCFW recursions some intermediate pieces can present non-local singularities, which are removed from the final result by mutual cancellations. On the other hand, gauge invariance is assured along each step. 
	}.

	\item[Analiticity] is the assumption that the amplitude is an analytic function of the kinematic variables. This eventually means that its singularity structure, since we have ruled out delta functions, is made of poles and branch cuts. Moreover, we are allowed to analytically continue the amplitude to complex momenta, in order to exploit the power of complex analysis and determine it from its singularities.
	
	\item[Unitarity] is essential to sensibly define a probability amplitude. For the $S$-matrix, as we have already seen, unitarity results in the condition~\eqref{unitarity}. This implies in particular that at the locus of a singularity, which corresponds to one or several of the involved particles going on-shell, the amplitude factorizes into sub-amplitudes with lower number of external legs and/or at lower perturbative order. It is clear that this factorization property is the crucial one to have recursion relations, and so we will discuss it a bit more in detail.

\end{description}

\begin{figure}[htb]
	\begin{minipage}{0.59\textwidth}
		\vspace{2mm}
		\flushright
		\includegraphics[width=\textwidth]{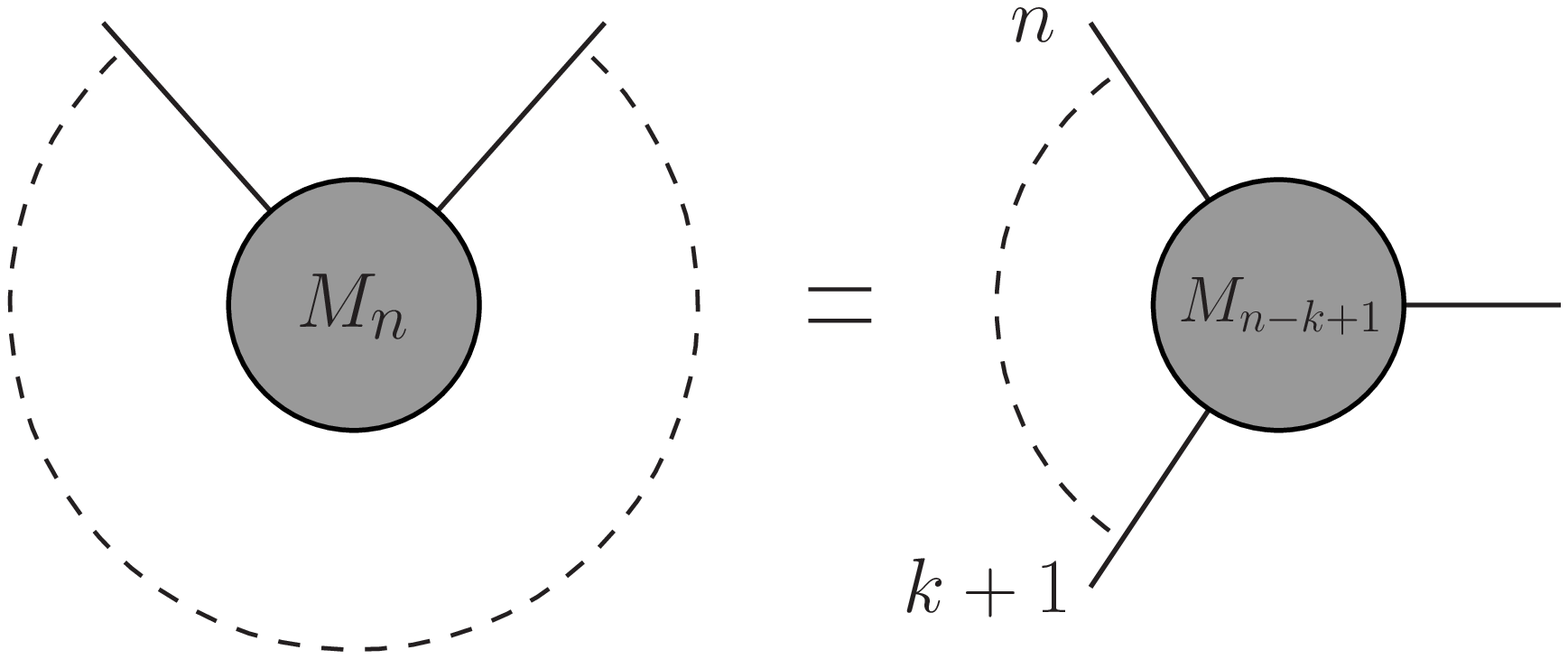}
	\end{minipage}
	\begin{minipage}[b]{0.12\textwidth}
		\centering
		\large$\displaystyle\frac{1}{\;\pp_{\KK}^2-m^2\;}$
	\end{minipage}
	\begin{minipage}{0.24\textwidth}
		\vspace{-0.5mm}
		\flushleft
		\includegraphics[width=\textwidth]{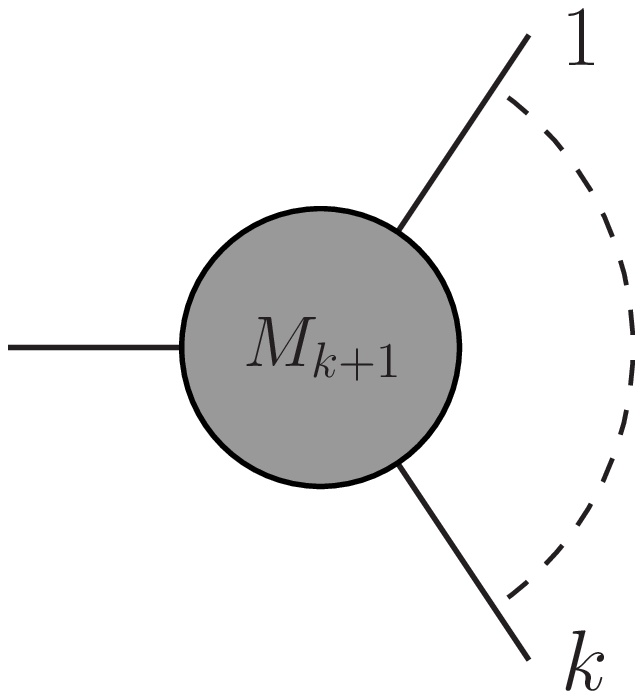}
	\end{minipage}
	\caption{Schematic representation of factorization of the amplitude around physical poles. The sum over different contributions is omitted.\label{BCFWfactor}}
\end{figure}

Consider an $n$-particles scattering process. Imagine that a subgroup~$\KK$ of the $n$ external momenta, containing $k$ of them, squares to one of the physical masses of the considered asymptotic states:
\begin{equation}	\label{pikappa}
\mathfrak{p}_{\KK}^2=\Big(\sum_{p_i\in\,\KK}\: p_i\Big)^{\!\!2}=m^2\ .
\end{equation}
This would corresponds to the production of an intermediate particle, which would then decay into the remaining $n-k$ particles, with the amplitude factorizing into two sub-amplitudes, with $k+1$ and $n-k+1$ external legs respectively, exchanging the intermediate particle (see fig.~\ref{BCFWfactor}). 
Of course we need $2\leq{k}\leq{n-2}$, for each of the sub-amplitudes in~\eqref{factor} to have at least three legs. 

Any intermediate state through which this factorization can occur is call \emph{factorization channel}. Such splitting of the total process into two sub-processes is not only possible, but infinitely more likely than all the particles interacting together at once. This infinitely greater probability is embodied by a simple pole singularity in the amplitude, located in momentum space where the on-shell condition of the intermediate particle is met. In formul\ae:
\begin{equation}	\label{factor}
M_n \;\;\sim\;\ \sum_{\KK}\ M_{n-k+1} \;\frac{1}{\phantom{|}\pp_{\KK}^2-m^2\phantom{|}}\; M_{k+1}\ ,
\end{equation}
where $M_n$ is at a given perturbative order, and $\pp_{\KK}$ is the one defined in~\eqref{pikappa}.\footnote{
	For a formal derivation of the factorization property from unitarity of the $S$-matrix, you can check Chapter~4 of the book ``The Analytic S-Matrix''~\cite{Eden:1966ab}, or Section~1.6 of Conde's lecture notes~\cite{Conde:2014mdv}.
} 

The formula~\eqref{factor} is potentially defining an \emph{on-shell} recursion relation, that is it can be used in the opposite direction (from right to left) to build an $n$-point amplitude from lower-point \emph{on-shell} information. We stress on the word \emph{on-shell} (both of the pieces that the amplitude factorizes in are physical, gauge-invariant, on-shell amplitudes), since there exist \emph{off-shell} recursion relations as well, where the amplitude is reconstructed recursively, but from off-shell information (Berends-Giele recursion relations~\cite{Berends:1987me} are an example).

Of course we should consider the possibility of more than one particle exchanged in the factorization channel, and so two or more internal propagators going simultaneously on-shell, yielding a branch cut, instead of a simple pole. This actually corresponds to loop contributions, whereas at tree-level we can only have simple poles. That is why we restrict to three level from now on, where we are completely able to establish recursion relations for the amplitude.

Now we have to provide an operational method to actually compute the contributions of the poles in~\eqref{factor}, in order to realize recursion relations. That will be achieved by using the power of complex analysis and Cauchy theorem, by virtue of the assumed analyticity of the $S$-matrix. \mbox{On-shell} recursion relations entail indeed complex deformation of some of the external momenta. Britto-Cachazo-Feng-Witten is a particular form of recursion relations, where only two (the minimal amount) of the external momenta are deformed. They were first discovered by Britto, Cachazo, and Feng~\cite{Britto:2004ap} in the context of one-loop Yang-Mills amplitudes, and then directly proven for generic tree-level amplitudes by the same three authors together with Witten~\cite{Britto:2005fq}.

Let us then consider a tree-level amplitude with $n$ external (real) momenta, and let us shift two of them, for $i=a,b$, in the following way
\begin{equation}	\label{shift}
p_a \;\longrightarrow\ p_a -z\,q \equiv \hat{p}_a\ , \qquad	p_b \;\longrightarrow\ p_b +z\,q \equiv \hat{p}_b\ , \qquad\ \text{with }\ z\in\mathbb{C}\ .
\end{equation}
Of course this kind of shift does not affect momentum conservation, $\sum_ip_i=0$. But we want to preserve `on-shellness' as well, then:
\begin{equation}
\left\{\begin{array}{l}
{\hat{p}_a}^2 = p_a^2 -2z\, q \cdot p_a +z^2\,q^2 = {p_a}^2	\vphantom{\frac{}{\big|}}\\
{\hat{p}_b}^2 = p_b^2 +2z\, q \cdot p_b +z^2\,q^2 = {p_b}^2
\end{array}\right.\quad\Leftrightarrow\quad \left\{
	\begin{array}{l}
	q^2 = 0	\vphantom{\frac{}{\big|}}\\
	q \cdot p_a = 0 = q \cdot p_b
	\end{array}\right.\ .
\end{equation}
So the shifting momentum $q$ must be light-like, and orthogonal to both $p_a$ and $p_b$. In four or higher dimensions, such a $q$ always exists, if we allow it to be complex.

Then the amplitude, expressed in terms of these shifted momenta, gets a dependency on the complex variable~$z$. It is by construction an holomorphic function of $z$, and it matches for $z=0$ the original amplitude. Then, moving away from the origin in the $z$-complex-plane, we will intercept the singular points, eventually corresponding to physical poles. The gain of deforming momenta is indeed that of translating the physical poles in the kinetic variables into poles in the unique holomorphic variable $z$.  Then we can use Cauchy theorem on the `shifted' amplitude~$\hat{M}^{(a,b)}_n(z)$, and state
\begin{equation}
R_n^{\infty}=\sum_{z_{\II}\neq0}\,\underset{\;z=z_{\II}}{\mathrm{Res}}\bigg[\frac{\hat{M}^{(a,b)}_n\!(z)}{z}\bigg]\ +\ \hat{M}^{(a,b)}_n(0)\ ,
\end{equation}
where $z_{\II}$ are the locations of the poles, $R_n^\infty$ represents the residue at complex infinity, and we have singled out from the sum the residue in zero, which is actually the physical, non-shifted amplitude that we want to determine. 

The residues at finite $z$ can be determined in a general fashion, whereas the residue at infinity can be an issue. In some specific cases it can be proven to be zero, or explicitly computed, either resorting to Lagrangian-based arguments, or to other principles. More in general, the existence or not of the residue at infinity depends on the pair of momenta, $p_a$ and $p_b$, that we decide to deform. So, as we will see, it may occur that for some particular shifts the residue at infinity vanishes, letting us to safely apply the recursion formula, whereas for some other shifts it does not vanish. In any case, we have to deal with this issue, if we want to apply these techniques.

Let us assume from now on that we are in a case where the residue at infinity vanishes for the chosen shift~(a,b). Then any pole $z_{\II}$ would correspond to a factorization channel, and so to a partition of the external particles into $\II$ and $\bar{\II}$ (complement of~$\II$), and from the factorization property~\eqref{factor} we can write
\begin{equation}	\label{BCFWsplit}
M_n=\hat{M}^{(a,b)}_n(0)=-\sum_{z_{\II}\neq0}\,\underset{\;z=z_{\II}}{\mathrm{Res}}\bigg[\frac{\hat{M}^{(a,b)}_n\!(z)}{z}\bigg]= \sum_{z_{\II}\neq0}\,\hat{M}_{\bar{\II}}(z_{\II})\;\frac{1}{\:\mathfrak{p}_{\II}^2-m^2\:}\;\hat{M}_{\II}(z_{\II})\ .
\end{equation}
Besides the sum over the poles, that is over specific partitions~$\II$ of the external particles, we have also to take a sum over all possible internal states (different masses~$m$, helicities, spins, etc...), which we have omitted here. Of course, the partition~$\II$ must contain at least two momenta and at most $n-2$ of them, since the sub-amplitudes cannot have less than three legs.

In order to be able to explicitly write down~\eqref{BCFWsplit}, we need to determine the locations of the poles. To do so, we require the shifted internal momentum to go on-shell at the location of the pole. Of course, for any given shift~$(a,b)$, only partitions where the shifted momenta $\hat{p}_a$ and $\hat{p}_b$ are on opposite sides give a shifted internal momentum, thus contributing to the sum. So let us call $\AA$ a subset of external momenta containing $p_a$, and $\BB$ the complementary subset, which is containing $p_b$. Then the internal momentum will be given by $\pp_{\AA}=\sum_{p_i\in\AA} p_i$, which will inherit the same shift~\eqref{shift} as~$p_a$: $\hat{\pp}_{\AA}=\pp_{\AA}-zq$. The location of the pole~$z_{\AA}$ is the value of $z$ such that the shifted momentum~$\hat{\pp}_{\AA}$ goes on-shell, that is:
\begin{equation}	\label{locus}
0=\hat{\pp}_{\AA}^2-m^2 = 
	\big(\pp_{\AA}^2-m^2\big)\bigg(1-z\,\frac{2q\cdot\pp_{\AA}}{\:\pp_{\AA}^2-m^2\,}\bigg) = 
		\big(\pp_{\AA}^2-m^2\big)\bigg(1-\frac{z}{z_{\AA}}\bigg) 
			\;\Rightarrow\;
				z_{\AA}=\frac{\pp_{\AA}^2-m^2}{2q\cdot\pp_{\AA}}\ .
\end{equation}

Then we can finally rewrite~\eqref{BCFWsplit} as
\begin{equation}	\label{BCFWsum}
M_n=\sum_{\AA}\,\hat{M}_{\BB}(z_{\AA})\;\frac{1}{\:\pp_{\AA}^2-m^2\:}\;\hat{M}_{\AA}(z_{\AA})\ ,
\end{equation}
where $z_A$ is given by~\eqref{locus}, and again we are omitting the sum over different internal physical states.

We have thus a very general picture of how on-shell recursion relation can be derived for tree-level amplitudes. We stress that, whereas the rest of these notes is firmly grounded in four dimensions, the derivation depicted here does not rely on anything specific to four dimensions, and it holds indeed in higher dimensions as well\footnote{
	In lower dimensions there is the issue that the shifting momentum $q$ with the required properties does not exist. However, in some cases recursion relations can be generalized to three dimensions, as for instance for Chern-Simons theories with matter~\cite{Gang:2010gy}.
}. Furthermore, our discussion is valid for massive particles as well as for massless ones, even if the original BCFW papers~\cite{Britto:2004ap,Britto:2005fq} were dealing only with massless Yang-Mills theory. The extension of BCFW recursion to massive particles is due to Badger et al.~\cite{Badger:2005zh,Badger:2005jv}. At the same time, the amplitudes and the involved momenta could be equivalently expressed in spinor-helicity formalism as well as in four-vector language. However, in particular for massless particles, since we are dealing with complexified and on-shell momenta, the spinor-helicity formalism is the ideal tool for expressing on-shell recursion relations; and indeed it was used in the original papers~\cite{Britto:2004ap,Britto:2005fq}.

So, before moving to some practical applications of BCFW recursions (for massless particles), we briefly reformulate the BCFW shift in spinor-helicity language.

\subsubsection*{BCFW shift in spinor-helicity formalism}

Let us consider the shift~\eqref{shift} when both $p_a$ and~$p_b$ are light-like. Then we can write $p_a=\lambda_a\tilde{\lambda}_a$, and~$p_b=\lambda_b\tilde{\lambda}_b$. The shifting momentum~$q$ has to be light-like anyway, so we also write $q=\mu\tilde{\mu}$. Then the BCFW shift~\eqref{shift} rephrases as
\begin{equation}
\lambda_a\tilde{\lambda}_a \;\longrightarrow\ \lambda_a\tilde{\lambda}_a -z\,\mu\tilde{\mu}\ , \qquad 	\lambda_b\tilde{\lambda}_b \;\longrightarrow\ \lambda_b\tilde{\lambda}_b +z\,\mu\tilde{\mu}\ ,
\end{equation}
and the additional orthogonality conditions for~$q$ read
\begin{equation}
\ket{\mu}{\lambda_a}\bra{\tilde{\lambda}_a}{\tilde{\mu}} = 0 = \ket{\mu}{\lambda_b}\bra{\tilde{\lambda}_b}{\tilde{\mu}}.
\end{equation}
There are two distinct solutions satisfying such conditions, that is either $q=\lambda_a\tilde{\lambda}_b$, or~$q=\lambda_b\tilde{\lambda}_a$. Notice that for both solutions only one spinor for $p_a$ and only one spinor for $p_b$ are shifted, namely:
\begin{equation}	\label{masslessshift}
q=\lambda_a\tilde{\lambda}_b \;\Rightarrow\ \left\{ 
	\begin{array}{l}
		\tilde{\lambda}_a \;\rightarrow\ \tilde{\lambda}_a -z\,\tilde{\lambda}_b	\vphantom{\frac{}{\big|}}\\
		\lambda_b \;\rightarrow\ \lambda_b +z\,\lambda_a
	\end{array}\right. \ ; \qquad 
q=\lambda_b\tilde{\lambda}_a \;\Rightarrow\ \left\{ 
	\begin{array}{l}
		\lambda_a \;\rightarrow\ \lambda_a -z\,\lambda_b	\vphantom{\frac{}{\big|}}\\
		\tilde{\lambda}_b \;\rightarrow\ \tilde{\lambda}_b +z\,\tilde{\lambda}_a
	\end{array}\right. \ .
\end{equation}
The first option is conventionally referred to as~$[a,b\rangle$-shift, while the second one as $\langle a,b]$-shift.

We are now ready to apply these techniques to build up tree-level massless amplitudes.

\subsection{Parke-Taylor formula}

Parke-Taylor formula is a stunning result for $n$-point tree-level gluon amplitudes, which was `empirically' inferred by Parke and Taylor in 1986~\cite{Parke:1986gb}. It is a formula for \emph{maximally helicity violating}~(MHV) $n$-gluon tree-level amplitudes in Yang-Mills theory. MHV means that all gluons have the same helicity, except for two of them. It is maximally helicity violating, since actually amplitudes where all gluons have the same helicity or at most one has different helicity both vanish for any number of external particles:
\begin{align}
&	A_n(\pm,\ldots,\pm)=0 \ , 		\label{MMMHV} \\
&	A_n(\mp,\pm,\ldots,\pm)=0 \ ,	\label{MMHV}
\end{align}
where we have used the letter $A$, rather than $M$ to specifically indicate a \emph{tree-level} amplitude, and the same we will do from now on. Moreover, we are here considering \emph{color-ordered} gluon amplitudes, which means that the color structure, coming from traces of the generators of the non-abelian gauge group, has been singled out, yielding an amplitude where the order of the particle is fixed, yet still enjoying a symmetry under cyclic permutations of the external legs. It is thanks to this cyclic invariance that we have always the right to move the gluon of different helicity to the first position. 

These results are recovered by Feynman graph calculations after cancellation of various (gauge dependent) terms. They can be proven by Lagrangian techniques based on Lorentz structures (see for instance Section 2.7 of~\cite{Elvang:2013cua}). At loop-level, these amplitudes are not vanishing anymore.

The first non trivial tree-level amplitudes are the MHV ones, where two gluons have different helicities from all the others, and Parke and Taylor realized that, even with increasing number of external legs, they keep a very simple form, \emph{i.e.}:
\begin{equation}	\label{ParkeTaylor}
\begin{aligned}
A_n(\ldots,i^-,\ldots,j^-,\ldots) &=
	\bfrac{{\ket{i}{j}}^4}{\ket12\ket23\cdots\ket{n-1}{n}\ket{n}{1}}\ ,	\\
A_n(\ldots,i^+,\ldots,j^+,\ldots) &=
	\bfrac{{\bra{j}{i}}^4}{\bra{1}{n}\bra{n}{n-1}\cdots\bra32\bra21}\ ,
\end{aligned}
\end{equation}
where we have omitted the powers of the coupling constant for the sake of neatness. This very simple results is again coming out of precise cancellations among different Feynman diagrams, whose number dramatically increases with increasing number of external legs~$n$:\footnote{
	See Appendix~A of~\cite{Kleiss:1988ne}, if you really want to check it...}
\begin{center}
\renewcommand{\arraystretch}{1.4}
\begin{tabular}{|R{2.5cm}||C{8mm}|C{8mm}|C{8mm}|C{8mm}|C{8mm}|C{8mm}|C{18mm}|}
\hline
$n$\hspace*{1ex} 				& 3	& 4	& 5		& 6		& 7		& $\cdots$ 	& 10 	\\
\hline
\# of diagrams\hspace*{1ex}	& 1 & 4	& 25	& 220	& 2485	& $\cdots$	& $10^{\cdot}525^{\cdot}900$	\\
\hline
\end{tabular}
\end{center}

The formula~\eqref{ParkeTaylor}, guessed by Parke and Taylor, was first proven by Berends and Giele~\cite{Berends:1987me} through their \emph{off-shell} recursion relations. We will show here the inductive proof~\cite{Britto:2005fq} based on BCFW \emph{on-shell} recursion relations.

The starting amplitude is the three-point one, which we can write, for two negative helicities and two positive helicities respectively, from the expressions~(\ref{M3H}-\ref{M3A})\footnote{
	The reader can notice the all-plus or all-minus amplitudes can be non-zero from~(\ref{M3H}-\ref{M3A}). However, they would have a coupling of different dimensions with respect to the coupling of the amplitudes~(\ref{PT3pH}-\ref{PT3pA}), corresponding thus to a different kind of cubic interaction, if considered as tree-level vertices. As we have already underlined, all-plus and all-minus amplitudes vanish only at tree-level in Yang-Mills theory, but they can be non trivial at loop-level. Expressions~(\ref{M3H}-\ref{M3A}) are non-perturbative, so they of course take into account beyond tree-level possibilities.
}. Again omitting the coupling constant, we have
\begin{align}
A_3(-,-,+)\equiv M_H^{-1,-1,+1} &= \frac{{\ket12}^4}{\ket12\ket23\ket31}\ , 	\label{PT3pH}\\
	A_3(+,+,-)\equiv M_A^{+1,+1,-1} &= \frac{{\bra21}^4}{\bra21\bra13\bra32}\ ,	\label{PT3pA}
\end{align}
which indeed match the Parke-Taylor expressions~\eqref{ParkeTaylor} for $n=3$.

So, we already have that Parke-Taylor formula holds for $n=3$. Now we want to use BCFW recursion to prove it for arbitrary~$n$. We assume that the formula holds for $n-1$, and we will obtain the formula for $n$ external legs. Another ingredient that we will use is the fact that all the amplitudes with at most one different helicity vanish~(\ref{MMMHV}-\ref{MMHV}). We postpone the proof of that at the end of this section, for the sake of readability. 

We will show the computation for the `mostly-plus' MHV amplitude, $M_n^{tree}(-,-,\ldots)$, the computation for the `mostly-minus' being completely identical. First of all we have to assure that there exists a shift that makes the residue at infinity vanish. As it was shown already in the original BCFW paper~\cite{Britto:2005fq}\footnote{
	The argument of~\cite{Britto:2005fq} is based on Feynman rules. In~\cite{Schuster:2008nh,He:2008nj} the large $z$ behaviors under different shifts for four-dimensional Yang-Mills theory have been determined also by non-Lagrangian approaches.
}, this is the case for gluon amplitudes with the shifts $\scriptstyle[-,-\rangle$ $({\scriptstyle\langle-,-]})$, $\scriptstyle[-,+\rangle$ $({\scriptstyle\langle+,-]})$, $\scriptstyle[+,+\rangle$ $({\scriptstyle\langle+,+]})$. On the contrary, the term at infinity does not vanish for the shift $\scriptstyle[+,-\rangle$ $({\scriptstyle\langle-,+]})$. As an indication that something is wrong with these latter combinations, we can check \emph{a fortiori} on the results~\eqref{ParkeTaylor} that they explode for $z\rightarrow\infty$ precisely when we shift the `non-tilded' spinor of a negative-helicity leg \emph{and} the `tilded' spinor of a positive-helicity leg; on the contrary, under all other shifts, we find a fall-off as $z^{-1}$ or faster.

We choose the valid shift~$\langle n^+,1^-]$, that is, from eq.s~\eqref{masslessshift},
\begin{equation}	\label{n1shift}
\lambda_n \;\rightarrow\ \lambda_n -z\,\lambda_1\ , \qquad
	\tilde{\lambda}_1 \;\rightarrow\ \tilde{\lambda}_1 +z\,\tilde{\lambda}_n\ ,
\end{equation}
which will turn out to make the computation particularly simple. Then we consider the factorization formula~\eqref{BCFWsplit}: since the amplitude factorizes around poles in $z$, then the shifted legs $\hat{1}$ and $\hat{n}$ have to appear in opposite sub-amplitudes, otherwise the internal momentum would not be shifted. Moreover, since we are considering color-ordered amplitudes, the position of the external legs cannot shuffle. Thus, we write
\begin{equation*}
A_n(1^-,2^-,\ldots) = \sum_{k=3}^{n-2}\sum_{\ h_k=\pm} A_{n-k+1}({k+1}^+,\ldots,\hat{n}^+,\hat{\mathfrak{p}}_k^{h_k}) \:\frac{1}{\,\mathfrak{p}_k^2\,}\: A_{k+1}(\hat{1}^-,2^-,\ldots,k^+,-\hat{\mathfrak{p}}_k^{-h_k})\ ,
\end{equation*}
where $\mathfrak{p}_k=p_1+\cdots+p_k=-(p_{k+1}+\cdots+p_n)$. You see that the exchanged momentum~$\mathfrak{p}_k$ has to appear with opposite sign and helicity in the two sub-amplitudes, since it is incoming on one side whereas is outgoing on the other side.
\begin{figure}[hb]
\centering
	\begin{minipage}{0.4\textwidth}
	\vspace{-3.2mm}
	\flushright
	\includegraphics[width=0.8\textwidth]{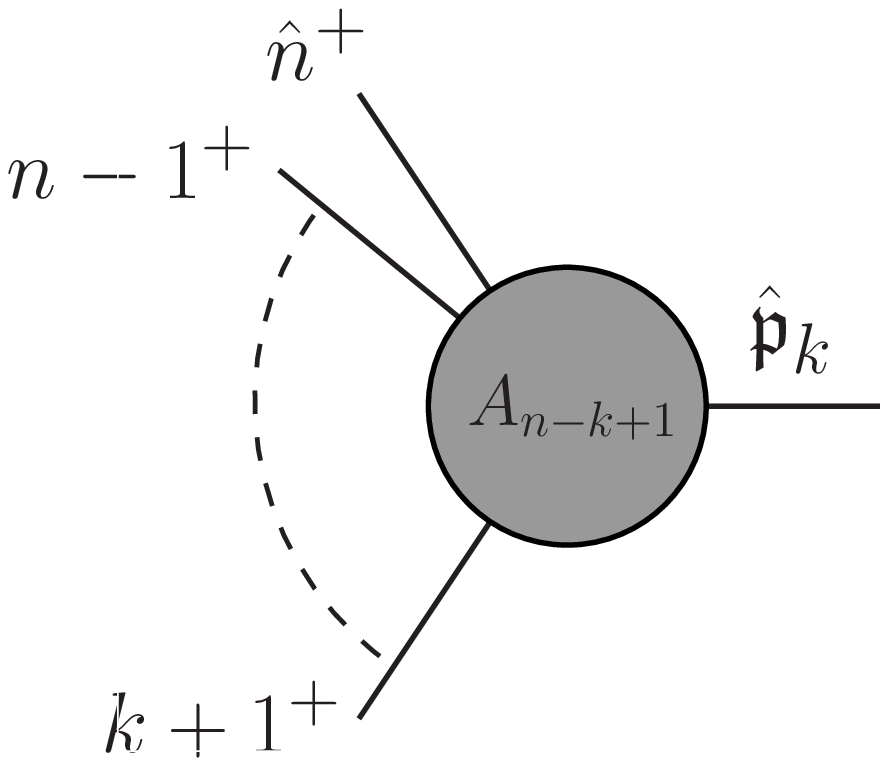}
	\end{minipage}
		\begin{minipage}[b]{0.07\textwidth}
		\centering
		\large$\displaystyle\frac{1}{\;\pp_{k}^2\;}$
		\end{minipage}
			\begin{minipage}{0.4\textwidth}
			\vspace{-3.2mm}
			\flushleft
			\includegraphics[width=0.69\textwidth]{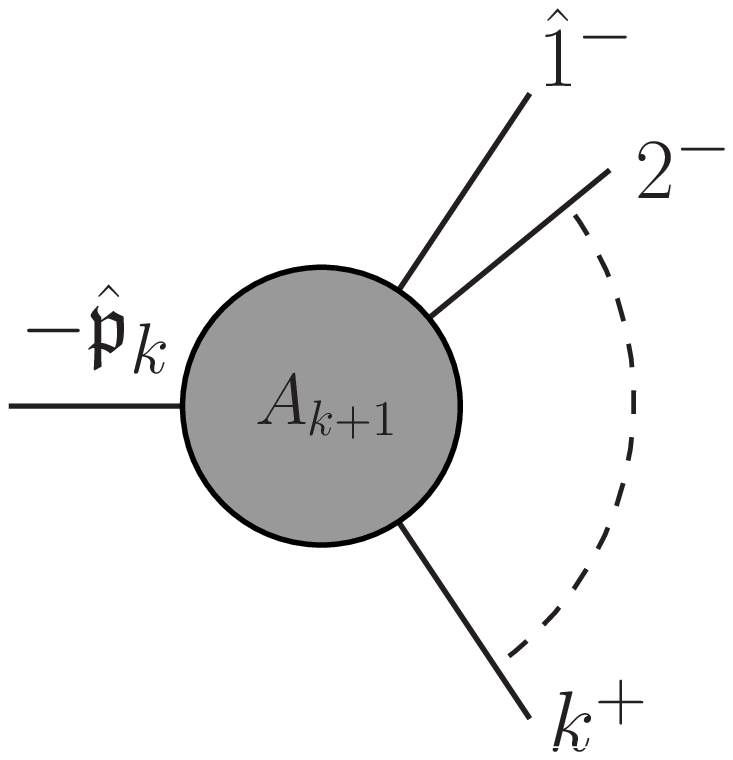}
			\end{minipage}
\caption{BCFW contribution for the $n$-point Parke-Taylor formula.\label{figPT}}
\end{figure}

Then we notice that the sub-amplitude on the left side can have at most one negative helicity, so for~(\ref{MMMHV}-\ref{MMHV}) it vanishes for any number of external legs, except three. The three-point amplitude for all positive helicities is zero as well, so the only non-vanishing contribution is for $k=n-2$ and $h_k=-1$, that is
\begin{equation}
A_n(1^-,2^-,\ldots) = 
	A_{3}({n-1}^+,\hat{n}^+,-\hat{\mathfrak{p}}_n^-)
		\;\frac{1}{\,\mathfrak{p}_n^2\,}\;
			A_{n-1}(\hat{1}^-,2^-,\ldots,{n-2}^+,+\hat{\mathfrak{p}}_n^{+})\ ,
\end{equation}
where $\mathfrak{p}_n=p_n+p_{n-1}$. You see that the $n$-point Parke-Taylor formula is indeed recovered from the $n-1$~version, together with the basic three-point block.

Inferring the expression for the three-point amplitudes from~\eqref{PT3pA}, and using the Parke-Taylor formula~\eqref{ParkeTaylor} for $n-1$, we obtain
\begin{equation}	\label{M3Mn-1}
A_n(1^-,2^-,\ldots) = \bfrac{{\bra{n}{n-1}}^3}{\bra{n-1}{\hat{\mathfrak{p}}_n}\bra{\hat{\mathfrak{p}}_n}{n}}\
	\frac{1}{\:\mathfrak{p}_n^2\:}\
		\bfrac{{\ket{1}{2}}^4}{\ket{1}{2}\cdots\ket{n-2}{-\hat{\mathfrak{p}}_n}\ket{-\hat{\mathfrak{p}}_n}{1}}\ ,
\end{equation}
where we have already used $\hat{n}]\equiv n]$ and $\hat{1}\rangle\equiv1\rangle$, since our shift~\eqref{n1shift} affects $\lambda_n$ and $\tilde{\lambda_{1}}$, and not $\tilde{\lambda}_n$ and $\lambda_{1}$. A technical point would be the choice of sign in expressing~$-\hat{\mathfrak{p}}_n=|-\hat{\mathfrak{p}}_n\rangle[-\hat{\mathfrak{p}}_n|=-|\hat{\mathfrak{p}}_n\rangle[\hat{\mathfrak{p}}_n|$ in terms of spinors. The $\mathbb{Z}_2$ ambiguity in the definition allows us to attribute the minus sign either to the `angle' spinor $(|-\hat{\mathfrak{p}}_n\rangle\equiv-|\hat{\mathfrak{p}}_n\rangle,\ [-\hat{\mathfrak{p}}_n|\equiv[\hat{\mathfrak{p}}_n|)$, or to the `square' spinor. However, it makes no difference, since $|-\hat{\mathfrak{p}}_n\rangle$ appears an even number of times in~\eqref{M3Mn-1}.

Then, we have
\begin{eqnarray*}
	\mathfrak{p}_n^2 &=& 2p_n\cdot p_{n-1} = \ket{n}{n-1}\bra{n-1}{n}\ , \\
	\ket{n-2}{\hat{\mathfrak{p}}_n}\bra{\hat{\mathfrak{p}}_n}{n} &=&
	\Pgen{n-2}{\hat{\mathfrak{p}}_n}{n} = \Pgen{n-2}{\hat{p}_n+p_{n-1}}{n} = \ket{n-2}{n-1}\bra{n-1}{n}\ ,	\\
	\ket{\hat{\mathfrak{p}}_n}{1}\bra{n-1}{\hat{\mathfrak{p}}_n} &=& 
	\Pgen{1}{\hat{\mathfrak{p}}_n}{n-1} = \ket{1}{\hat{n}}\bra{n}{n-1} = \ket{1}{n}\bra{n}{n-1}\ ,
\end{eqnarray*}
where we have used the definitions~(\ref{ket}--\ref{Pgen}), and in the last line the fact that $\ket{1}{\hat{n}}=\ket{1}{n}$, which comes straightforwardly from~\eqref{n1shift}.

Putting all this into~\eqref{M3Mn-1} we immediately get the desired result:
\begin{equation*}\label{Mn12}
A_n(1^-,2^-,\ldots) = \bfrac{{\ket12}^4}{\ket12\cdots\ket{n}{1}}\ .
\end{equation*}
\smallskip

The simplicity and conciseness of this proof is one of the neatest demonstration of the power of BCFW recursion for tree-level amplitudes. The key element of such power is that the different contributions to BCFW recursion relation are made up of gauge-invariant objects: the on-shell sub-amplitudes and the exchanged propagators. The gauge redundancy of local Lagrangians leads to a proliferation of gauge dependent contributions, the Feynman diagrams. Gauge invariance is restored in the final result upon cancellations among the various gauge dependent terms. So, the gauge redundancy is somehow the price to pay for manifest locality, which we have in a local field theory. 

On the contrary, in on-shell factorization, locality is not manifest: it underlies in the fact that singularities come from poles of propagators. However, in more involved calculations than this one, where more than one contributions sum up, some additional `spurious' poles can arise, meaning poles which are not those of the physical factorization channels. These non-physical poles eventually cancel out among different contributions in the recursive formula, in an analogous way as gauge dependency is cleared up by sum of different Feynman diagrams. However, generally on-shell recursive techniques (where they apply) drastically reduce the number of contributions with respect to Feynman diagrams.

Moreover, we have noticed that, provided we can assure that we have no residue at infinity, we are free to choose the BCFW-shift among various possibilities. Of course, different shifts must yield the same result for a given amplitude, yet the number of contributions and the simplicity of the computation may depend on the chosen shift. So, it is often possible to choose a more favorable shift that optimizes the BCFW computation\footnote{
	For instance, we invite you to derive again the result~\eqref{Mn12} by the $[{n-1}^+,n^+\rangle$ or $[1^-,2^-\rangle$ shifts. You would encounter one difficulty more with respect to the computation presented here.
}.

On the other hand, this possibility of computing the same amplitude through different BCFW shifts can also be used to constrain the amplitude. Indeed, in some cases different shifts give different results for the same amplitude. Requiring the final answer to be unique thus gives some conditions on the considered amplitude, and so on the form of the relative interaction. It is the idea at the core of the \emph{four-particle test}, which we illustrate in the next section.
\medskip

Before moving to next section, let us show that starting form the three-point vertices~(\ref{PT3pH}-\ref{PT3pA}) we cannot obtain non-vanishing amplitudes with at most one different helicity~(\ref{MMMHV}-\ref{MMHV}), which completes our proof of Parke-Taylor formula.

Quite straightforwardly, the four-amplitudes with all negative or all positive helicities cannot be generated from the considered vertices~(\ref{PT3pH}-\ref{PT3pA}), since, as we have seen, the intermediate particle must have opposite helicity on opposite sides. If $A_4(\mp,\mp,\mp,\mp)=0$, then recursively all other amplitudes with all equal helicities are zero as well.

In order to prove that also the amplitudes where only one helicity differs from the others are zero~\eqref{MMHV}, we have to show that certain BCFW contributions vanish. In particular, we will see that any three-point sub-amplitude of the form~(\ref{PT3pH}-\ref{PT3pA}) where one of the shifted spinors appears explicitly is zero. 

This can be intuitively understood recalling that in an on-shell \emph{`holomorphic'} three-point amplitude~\eqref{M3H} the relative \emph{`tilded'} spinors are all proportional (all the \emph{square-products} vanishing), while in an on-shell \emph{`anti-holomorphic'} three-point amplitude~\eqref{M3A} the relative \emph{`non-tilded'} spinors are all proportional (all the \emph{angle-products} vanishing). We have already seen in the derivation of the Parke-Taylor formula that the shifted momenta have to be on different sides of the factorization in order to contribute: in such case we use the shifting variable~$z$ to put the exchanged momentum on-shell, so getting on-shell amplitudes on both sides of the factorization. With the BCFW shift, we shift \emph{only one} `non-tilded' spinor and \emph{only one} `tilded' spinor. The shift of the `tilded' (`non-tilded') spinor makes all the square-products (angle-products) vanish at the location of the pole, so that the `holomorphic' (`anti-holomorphic') three-point sub-amplitude is on-shell and non-zero, whereas the `anti-holomorphic' (`holomorphic') sub-amplitude would contain the shifted `tilded' (`non-tilded') spinor and be zero.

Let us see this in a practical case, which will be a bit tedious, but useful to be convinced once for all, and then automatically discard these kind of contributions in any future computation. Let us thus consider a BCFW contribution of the kind
\begin{equation}	\label{zerocontrib}
A_3(-\hat{\mathfrak{p}}_{ij}^-,i^{\,-},\hatj^{\,+}) \;\frac{1}{{\mathfrak{p}_{ij}}^2}\; \cdots\ =
\bfrac{{\ket{\hat{\mathfrak{p}}_{ij}}{i}}^3}{\ket{i}{\hatj}\ket{\hatj}{\hat{\mathfrak{p}}_{ij}}} \;\frac{1}{{\mathfrak{p}_{ij}}^2}\; \cdots\ ,
\end{equation}
where $\hat{\mathfrak{p}}_{ij}=p_i+\hat{p}_j$, and the dots stands for the other on-shell sub-amplitude which completes the factorization. We are performing a $\langle j^+,k^{h_k}]$-shift (which is a `safe' shift, independently of the value of~$h_k$!), with $p_k$ being of course in the sub-amplitude in the dots, and so we are shifting the `non-tilded' spinor of $p_j$:
\begin{equation}
\lambda_j \;\rightarrow\ \lambda_j -z\,\lambda_k\ .
\end{equation}
We want to prove that the three-point sub-amplitude in~\eqref{zerocontrib}, which explicitly exhibits the shifted spinor~$\lambda_{j}$, is made of spinor products that all vanish at the location of the pole, and so vanishes as well (three powers in the numerator dominate over two powers in the denominator).

Requiring as usual the intermediate momentum to be on-shell, we find the location of the pole~$z_{ij}$,
\begin{equation}	\label{polezij}
0=\hat{\mathfrak{p}}_{ij}^2=\ket{i}{\hatj}\bra{j}{i}=\bra{j}{i}\left(\ket{i}{j}-z_{ij}\ket{i}{k}\right) \quad\Leftrightarrow\quad
z_{ij}=\bfrac{\ket{i}{j}}{\ket{i}{k}}\ .
\end{equation}
Therefore at the location of the pole the angle-product~$\ket{i}{\hatj}$ goes to zero. Then we have
\begin{equation}	\label{lambdapij}
\begin{aligned}
\!\!\!
\hat{\mathfrak{p}}_{ij}=
	\lambda_{i}\tilde{\lambda}_i+\left(\lambda_{j}-z_{ij}\,\lambda_k\right) \tilde{\lambda}_j =
		\lambda_{i}\tilde{\lambda}_i-\frac{\ket{k}{i}\,\lambda_{j}+\ket{i}{j}\,\lambda_k}{\ket{i}{k}}\,\tilde{\lambda}_j =
			\lambda_{i}\Big(\tilde{\lambda}_i+\frac{\ket{j}{k}}{\ket{i}{k}}\,\tilde{\lambda}_j\Big) \qquad &\\
\Rightarrow\
|\hat{\mathfrak{p}}_{ij}\rangle=\lambda_i\ , \quad
[\hat{\mathfrak{p}}_{ij}|= \tilde{\lambda}_i+\frac{\ket{j}{k}}{\ket{i}{k}} \tilde{\lambda}_j\ ,&
\end{aligned}
\end{equation}
where we have used the Schouten identity~\eqref{schouten} in the last step of the first line. Thus, we have that $\ket{\hatj}{\hat{\mathfrak{p}}_{ij}}=\ket{\hatj}{i}$, which goes to zero at the location of the pole~\eqref{polezij}. To deal with the numerator in~\eqref{zerocontrib}, we multiply and divide\footnote{
	Of course, we have to worry about whether we are multiplying and dividing for something vanishing. Using the value of $[\hat{\mathfrak{p}}_{ij}|$ in~\eqref{lambdapij}, it can be immediately checked that $\bra{i}{\hat{\mathfrak{p}}_{ij}}$ is not zero at the location of the pole.
} for $\bra{i}{\hat{\mathfrak{p}}_{ij}}^3$, and use
\begin{equation*}
\bra{i}{\hat{\mathfrak{p}}_{ij}}\ket{\hat{\mathfrak{p}}_{ij}}{i}=\Pgen{i}{\hat{\mathfrak{p}}_{ij}}{i} = \ket{i}{\hatj}\bra{j}{i}\ ,
\end{equation*}
which is also proportional to the vanishing angle-product~$\ket{i}{\hatj}$.

Thus, as announced, all the spinor products of the three-point sub-amplitude in~\eqref{zerocontrib} go to zero at the location of the pole as~$\ket{i}{\hatj}\propto z-z_{ij}$. In virtue of one power more at the numerator, the three-point sub-amplitude vanishes, and so the whole BCFW term~\eqref{zerocontrib}.

With identical procedure, it can be checked for all the cases where the \emph{`non-tilded'} shifted spinor explicitly appears in a `holomorphic' sub-amplitudeand for all the cases where the \emph{`tilded'} shifted spinor explicitly appears in a `anti-holomorphic' three-point sub-amplitude. We summarize these useful results, which will be helpful already in next section:
\begin{equation}	\label{vanish3p}
\begin{aligned}	
A_3(-\hat{\mathfrak{p}}_{ij}^-,i^{\,-},\hatj^{\,+}) = 
	A_3(-\hat{\mathfrak{p}}_{ij}^-,\hatj^{\,-},i^{+}) =
		A_3(i^{\,-},\hatj^{\,-},-\hat{\mathfrak{p}}_{ij}^+) &=0\ , 
\quad		\text{when $\lambda_j$ is shifted;} 	\vphantom{\frac{}{\big|}}\\	
A_3(-\hat{\mathfrak{p}}_{kl}^+,l^{+},\hat{k}^{-}) \,=
	A_3(-\hat{\mathfrak{p}}_{kl}^+,\hat{k}^{+},l^{-}) =\,
		A_3(\hat{k}^{+},l^{+},-\hat{\mathfrak{p}}_{kl}^-) &=0\ ,
\quad		\text{when $\tilde{\lambda}_k$ is shifted.}
\end{aligned}
\end{equation}

\begin{figure}[bt]
		\centering
		\includegraphics[width=\textwidth]{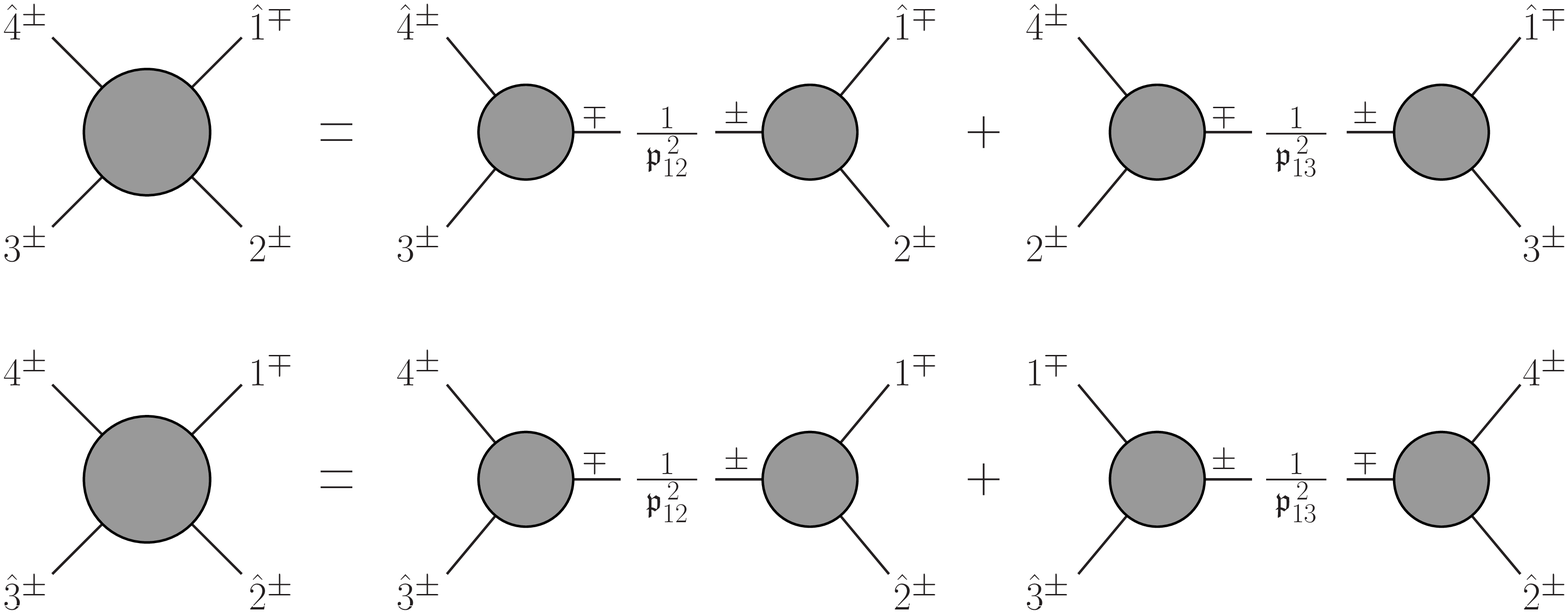}
\caption{BCFW formula of the four-point gluon amplitude with one different helicity, for the two possible shifts: $[-,+\rangle$ and~$[\pm,\pm\rangle$. \label{fig4pMMHV}}
\end{figure}

With these results we can now analyze the four-point amplitude where all helicities are equal but one. As you can see from figure~\ref{fig4pMMHV}, such four-point amplitudes necessarily factorizes into two three-point sub-amplitudes of the same kind, that is either both `holomorphic'~\eqref{PT3pH}, or both `anti-holomorphic'~\eqref{PT3pH}. In both cases, since the shifted momenta have to be on different sides of the factorization, one of the two sub-amplitudes contains the `non-tilded' shifted spinor, and the other one contains the `tilded' shifted spinor. So, either they are both `holomorphic' or both `anti-holomorphic', one of the two sub-amplitudes will explicitly exhibit one of the shifted spinors, and so it will vanishes. Thus, the whole four-point amplitude where all helicities but one are equal is zero. Then we straightforwardly obtain by induction that all higher-point amplitudes of the form~\eqref{MMHV} vanish as well.

\subsection{Four-particles test}

In principle, if we compute a given amplitude through BCFW with different shifts, we expect to obtain the same result. However, if we specify to some particular form of three-point interaction, different shifts can give different results for the same amplitude. This of course is not sensible, and so we require the two results to be equivalent, deriving some condition on the kind of interaction we are considering. Essentially, we demand the four-point amplitude to be compatible with the chosen form of three-point interaction. This requirement, very similar in spirit to what is done in the bootstrap paradigm, was dubbed `four-particle test' by Benincasa and Cachazo, as they introduced it in their work~\cite{Benincasa:2007xk}.

We present here as an example a particularly nice application which was worked out in that paper~\cite{Benincasa:2007xk}. We consider the cubic vertices of Yang-Mills gauge vector bosons (without color-ordering), that is vertices made of three spin-1 massless particles of various species (colors). The four-particle test in this case requires the coupling constants of such vertices to satisfy the Jacobi identity, letting emerge their connection with the structure constants of a non-abelian gauge group. This result is quite surprising, since in our basic initial ingredients (the existence of Lorentz-invariant three-point amplitudes of massless vector bosons of different kinds) there is no assumption of an underlying Lie algebra.

The three-point vertices we are going to consider are thus
\begin{equation}	\label{3pvertex}
A(i^-_a,j^-_b,k^+_c)=\kappa_{abc}\bfrac{{\ket{i}{j}}^3}{\ket{j}{k}\ket{k}{i}}\ , \qquad
	A(i^+_a,j^+_b,k^-_c)=\kappa_{abc}\bfrac{{\bra{j}{i}}^3}{\bra{i}{k}\bra{k}{j}}\ ,
\end{equation}
whose form is given by~(\ref{M3H}, \ref{M3A}), where we allow for different coupling constants depending on the species of the external particles. On the contrary, we take the coupling to be the same for both expressions in~\eqref{3pvertex}, that is for vertices with the same species of particles but opposite helicities. This corresponds to consider a parity-invariant interaction (sign inversion of the spatial part of momentum produces flipping of the helicity). Since the two vertices are connected by complex conjugation, we also have that the couplings are real. Moreover, these amplitudes flip sign when we exchange two particles, so the couplings must be completely antisymmetric in order not to violate the crossing symmetry (changing labels should not affect the amplitude). Finally, from the fact that the three-point amplitude has mass dimensions one, we have that the considered coupling is dimensionless, $[\kappa_{abc}]=0$.

We want to build four-point amplitudes out of this type of three-point interaction, using BCFW techniques. As we have seen in detail at the end of previous section, because of~\eqref{vanish3p}, the four-point amplitudes where at most one helicity differs from the others are all zero. So the only remaining possibility is the configuration with two positive and two negative helicities. Let us explicitly work out the case~$A_4(1_a^-,2_b^+,3_c^-,4_d^+)$, first using the~$[1^-,2^+\rangle$-shift, then the~$[1^-,4^+\rangle$-shift, and eventually requiring the two outcomes to be equal.

\begin{figure}[t]
	\centering
	\includegraphics[width=0.95\textwidth]{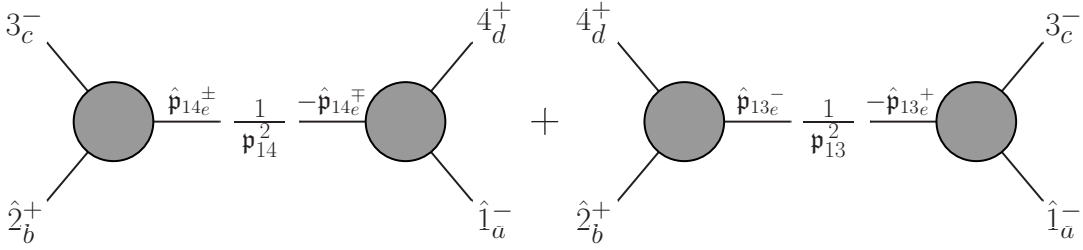}
	\caption{BCFW expression for $A_4\big(1_a^-,2_b^+,3_c^-,4_d^+\big)$ in the $[1,2\rangle$-shift. \label{fig4pTest}}
\end{figure}

For the the~$[1^-,2^+\rangle$-shift, we have the following two contributions (fig.~\ref{fig4pTest})\footnote{
	As displayed in figure~\ref{fig4pTest}, with the first contribution we could have considered also the one with flipped helicity for the intermediate momentum~$\hat{\mathfrak{p}}_{14}$: but this term is zero, again because it involves three-point amplitudes that explicitly contain the shifted variables,~$\tilde{\lambda}_1$ and~$\lambda_2$, which thus vanish at the location of the pole, as one of~\eqref{vanish3p}.
	}
\begin{align}
& A_4^{[1,2\rangle}\big(1_a^-,2_b^+,3_c^-,4_d^+\big) = 		\label{M4-12contrib} \\
&\qquad\quad =
	A\big({\hat{\mathfrak{p}}_{14}\vphantom{|}}^{\!+}_e,\hat{2}_b^+,3_c^-\big)\, \frac{1}{\mathfrak{p}_{14}^2}\, A\big(\hat{1}_a^-,-{\hat{\mathfrak{p}}_{14}\vphantom{|}}^{\!-}_e,4_d^+\big) +
		A\big(\hat{2}_b^+,4_d^+,{\hat{\mathfrak{p}}_{13}\vphantom{|}}^{\!-}_e\big)\, \frac{1}{\mathfrak{p}_{13}^2}\, A\big(3_c^-,\hat{1}_a^-,-{\hat{\mathfrak{p}}_{13}\vphantom{|}}^{\!+}_e\big) \nn
&\qquad\quad =
	\bfrac{\kappa_{dae}\kappa_{bce}}{\ket14\bra41}\, \bfrac{{\bra{2}{\hat{\mathfrak{p}}_{14}}}^3}{\bra{\hat{\mathfrak{p}}_{14}}{3}\bra32}
	\bfrac{{\ket{1}{\hat{\mathfrak{p}}_{14}}}^3}{\ket{\hat{\mathfrak{p}}_{14}}{4}\ket{4}{1}} +
		\bfrac{\kappa_{cae}\kappa_{bde}}{\ket13\bra31}\, \bfrac{{\bra42}^3}{\bra{2}{\hat{\mathfrak{p}}_{13}}\bra{\hat{\mathfrak{p}}_{13}}{4}}
		\bfrac{{\ket{3}{1}}^3}{\ket{1}{\hat{\mathfrak{p}}_{13}}\ket{\hat{\mathfrak{p}}_{13}}{3}}\ , \nonumber
\end{align}
with $\mathfrak{p}_{14}=p_1+p_4\equiv-p_2-p_3$, and~$\mathfrak{p}_{13}=p_1+p_3\equiv-p_2-p_4$.  The location of the poles is given by
\begin{align}
&	0=\hat{\mathfrak{p}}_{14}^2=\ket14\left(\bra41-z_{14}\bra42\right)\quad \Longleftrightarrow\quad z_{14}=\bfrac{\bra14}{\bra24}=-\bfrac{\ket23}{\ket13}\ ; \label{z14}\\
&	0=\hat{\mathfrak{p}}_{13}^2=\ket13\left(\bra31-z_{13}\bra32\right)\quad \Longleftrightarrow\quad z_{13}=\bfrac{\bra31}{\bra32}=-\bfrac{\ket24}{\ket14}\ .	\label{z13}
\end{align}
For the first term in~\eqref{M4-12contrib}, we then work out
\begin{align}
\ket{1}{\hat{\mathfrak{p}}_{14}}\bra{\hat{\mathfrak{p}}_{14}}{2} &= \ket14\bra42\ , \\
	\ket{4}{\hat{\mathfrak{p}}_{14}}\bra{\hat{\mathfrak{p}}_{14}}{3} &= \ket41\bra{\hat{1}}{3} = -\bfrac{\ket14}{\bra24}\big(\bra13\bra24+\bra32\bra14\big) = \bfrac{\ket14}{\bra24}\bra21\bra34\ ,
\end{align}
where for the second line we have used the expression of the location of the pole~$z_{14}$~\eqref{z14} and the Schouten identity~\eqref{schouten}; and
\begin{align}
\ket{1}{\hat{\mathfrak{p}}_{13}}\bra{\hat{\mathfrak{p}}_{13}}{4} &= \ket13\bra34\ , \\
	\ket{3}{\hat{\mathfrak{p}}_{13}}\bra{\hat{\mathfrak{p}}_{13}}{2} &= \ket31\bra{\hat{1}}{2} = \ket31\bra12\ ,
\end{align}
for the second term in \eqref{M4-12contrib}. We obtain so
\begin{equation*}
A_4^{[1,2\rangle}\big(1_a^-,2_b^+,3_c^-,4_d^+\big) = \bfrac{\kappa_{dae}\kappa_{bce}}{\ket14\bra41}\, \bfrac{\ket14{\bra24}^4}{\bra21\bra32\bra34} +
	\bfrac{\kappa_{cae}\kappa_{bde}}{\ket13\bra31}\, \bfrac{{\ket13}^2{\bra24}^3}{\ket13\bra21\bra34}\ ,
\end{equation*}
This expression can be simplified further thanks to momentum conservation, since
\begin{align*}
\ket14\bra24 &= -\Pgen{1}{p_4}{2}=\Pgen{1}{p_1+p_2+p_3}{2}=\ket13\bra32, \nn
\ket13\bra34 &=\Pgen{1}{p_3}{4}= -\ket12\bra24, \nn
\end{align*}
and it finally nicely reads
\begin{equation}	\label{M4-12}
A_4^{[1,2\rangle}\big(1_a^-,2_b^+,3_c^-,4_d^+\big) = -\frac{{\ket13}^2{\bra24}^2}{\sf s}\: \bigg(\frac{\kappa_{dae}\kappa_{bce}}{\sf t} +
\frac{\kappa_{cae}\kappa_{bde}}{\sf u}\bigg)\ ,
\end{equation}
where the standard definitions of the Mandelstam variables are employed, \emph{i.e.}: 
\begin{equation}
{\sf s}=(p_1+p_2)^2\ ,\quad {\sf t}=(p_1+p_4)^2\ ,\quad {\sf u}=(p_1+p_3)^2\ .
\end{equation}

We now compute the same amplitude through the $[1^-,4^+\rangle$-shift. Similarly to the previous case, we have the following two contributions
\begin{align*}
& A_4^{[1,4\rangle}\big(1_a^-,2_b^+,3_c^-,4_d^+\big) = \\
&\qquad\quad =
	A\big(\hat{4}^+_d,{\hat{\mathfrak{p}}_{12}\vphantom{|}}^{\!+}_e,3_c^-\big)\, \frac{1}{\mathfrak{p}_{12}^2}\, A\big(-{\hat{\mathfrak{p}}_{12}\vphantom{|}}^{\!-}_e,\hat{1}_a^-,2_b^+\big) +
		A\big(2_b^+,\hat{4}_d^+,{\hat{\mathfrak{p}}_{13}\vphantom{|}}^{\!-}_e\big)\, \frac{1}{\mathfrak{p}_{13}^2}\, A\big(\hat{1}_a^-,3_c^-,-{\hat{\mathfrak{p}}_{13}\vphantom{|}}^{\!+}_e\big) \\
&\qquad\quad =
	\bfrac{\kappa_{abe}\kappa_{cde}}{\ket12\bra21}\, \bfrac{{\bra{\hat{\mathfrak{p}}_{12}}{4}}^3}{\bra43\bra{3}{\hat{\mathfrak{p}}_{12}}} \bfrac{{\ket{\hat{\mathfrak{p}}_{12}}{1}}^3}{\ket12\ket{2}{\hat{\mathfrak{p}}_{12}}} +
		\bfrac{\kappa_{ace}\kappa_{bde}}{\ket13\bra31}\, \bfrac{{\bra42}^3}{\bra{2}{\hat{\mathfrak{p}}_{13}}\bra{\hat{\mathfrak{p}}_{13}}{4}} \bfrac{{\ket13}^3}{\ket{3}{\hat{\mathfrak{p}}_{13}}\ket{\hat{\mathfrak{p}}_{13}}{1}}\ .
\end{align*}
The location of the poles are now given by
\begin{align}
&	0=\hat{\mathfrak{p}}_{12}^2=\ket12\left(\bra21-z_{12}\bra24\right)\quad \Longleftrightarrow\quad z_{12}=\bfrac{\bra21}{\bra24}=-\bfrac{\ket34}{\ket31}\ ; \\
&	0=\hat{\mathfrak{p}}_{13}^2=\ket13\left(\bra31-\tilde{z}_{13}\bra34\right)\quad \Longleftrightarrow\quad \tilde{z}_{13}=\bfrac{\bra31}{\bra34}=-\bfrac{\ket42}{\ket12}\ .
\end{align}
With completely analogous manipulations as in the previous computation, we get
\begin{align}
A_4^{[1,4\rangle}\big(1_a^-,2_b^+,3_c^-,4_d^+\big) & = 
	\bfrac{\kappa_{abe}\kappa_{cde}}{\ket12\bra21}\, \bfrac{\ket12{\bra24}^4}{\bra14\bra43\bra32} + \bfrac{\kappa_{ace}\kappa_{bde}}{\ket13\bra31}\, \bfrac{{\ket13}^2{\bra24}^3}{\ket13\bra32\bra14} \nn
&	= 	\label{M4-14}
	-\frac{{\ket13}^2{\bra24}^2}{\sf t}\: \bigg(\frac{\kappa_{abe}\kappa_{cde}}{\sf s} + \frac{\kappa_{ace}\kappa_{bde}}{\sf u}\bigg)\ .
\end{align}
We have repeated the computation for completeness, but notice we could have obtained the expression~\eqref{M4-14} from~\eqref{M4-12}, by simply exchanging the labels 2 with 4 and $b$ with $d$.

Now we have nothing else to do than comparing the two different results for~$M_4$,~\eqref{M4-12} and~\eqref{M4-14}, and require them to coincide. This yields
\begin{align}
0 = A_4^{[1,2\rangle}-A_4^{[1,4\rangle} &=
	\frac{{\ket13}^2{\bra24}^2}{\sf st}\; \bigg(\kappa_{abe}\kappa_{cde} -\kappa_{dae}\kappa_{bce} +\frac{\kappa_{ace}\kappa_{bde}({\sf s+t})}{\sf u}\bigg)	\nn
&	=
	\frac{{\ket13}^2{\bra24}^2}{\sf st}\; \Big(\kappa_{abe}\kappa_{cde} +\kappa_{ade}\kappa_{bce} +\kappa_{ace}\kappa_{dbe}\Big)\ ,
\end{align}
where we have used the antisymmetric property of~$\kappa$, and the fact that ${\sf s+t+u}=0$ by momentum conservation. We have so retrieved the Jacobi identity for the coupling constants, as announced,
\begin{equation}
\kappa_{abe}\kappa_{cde} +\kappa_{ace}\kappa_{dbe} +\kappa_{ade}\kappa_{bce} =0\ .
\end{equation}
We know that from Yang-Mills theory the coupling constants of the gauge vector bosons are expressed as $g_{YM}\,f_{abc}$, where $f_{abc}$ are the structure constants of the Lie algebra of the underlying non-abelian gauge group, which indeed satisfy by definition the Jacobi identity.

With this charming result, which is a simple but surprisingly non-trivial test of BCFW techniques, we conclude our limited review of applications of BCFW. We hope that we have given the reader the starting tools for continuing the exploration of such a promising technique.

\section*{Epilogue}

We have arrived to the conclusion of our short travel through the world of scattering amplitudes. We hope these notes be a self-consistent story, which can be read in a whole, from the beginning to the end. At same time we wish them to be easily accessible at any point by the reader who is looking for a specific information.\footnote{
	We would be grateful to the reader for any commentary or remark, to be addressed to the \href{mailto:andrea.marzolla@ulb.ac.be}{author}.
} Before bidding farewell to the reader, we shortly summarize the outlines of our discussion.

After having recalled the basic properties of a scattering amplitude deriving from a consistent $S$-matrix theory, we have assumed Poincar\'e invariance as the symmetry of the spacetime. The representation theory of Poincar\'e group allowed us to define the asymptotic states corresponding to fundamental particles participating in the scattering process, and provided some constraints for the amplitude. These constraints are coming from the Little Group transformation properties, which are distinct for massless and massive particles. Nonetheless, for the simplest case of the three-point amplitude, the Little Group equations, together with momentum conservation and on-shell conditions, are able to completely fix the kinematic dependency of the amplitude, either the external legs are massless or massive. Some of these results vanish for real kinematics, but some others correspond to physical processes, constituting thus \emph{non-perturbative} expressions.

Then we have gone beyond the three-point case, thanks to BCFW recursion relations, which on the other hand are based on the validity of a perturbative expansion. We have used them at tree-level to prove by induction the Parke-Taylor formula, and to derive the Jacobi identity for non-abelian Yang-Mills couplings.

As already mentioned, the BCFW shift has been generalized to massive particles~\cite{Badger:2005zh,Badger:2005jv}, so a natural continuation of the subject presented here would be to use the massive three-point amplitudes~(\ref{sol1m}, \ref{sol2m}, \ref{sol3m}) as building blocks to construct higher-point \emph{massive} amplitudes. 

The reader may wonder about the extension of recursive techniques to loop-level. Actually, it is a very though open question, due to the much more involved singularity structure of loop level. First of all, loop-level recursive techniques apply to the integrand rather than the whole amplitude. Then, several on-shell analytic methods for loop integrands exist (see~\cite{Britto:2010xq} for a relatively recent review), but the direct extension of BCFW to loop-level presents some obstructions (well illustrated in Section~7 of~\cite{Elvang:2013cua}). In the special case of the planar limit of $\mathcal{N}=4$ super-Yang-Mills, such obstructions have been overcome, and full recursive formul{\ae} for all-loop amplitudes have been derived, based on momentum-twistor duality and the positive Grassmannian~\cite{ArkaniHamed:2010kv}.

In conclusion, we hope to have given a flavor of the fact that on-shell methods, non-perturbative as perturbative ones, if they are far from replacing local quantum field theory, as it was maybe the intention of the original $S$-matrix program, constitute at least a practically useful and theoretically enlightening alternative, which complements and extends Lagrangian-based techniques. 
\bigskip

\subsection*{Aknowledgements}
\noindent
Following chronological order, let me thank Gabriele Travaglini, since it is by his lectures at Lisbon \emph{Summer School on String Theory and Holography} in July 2014 that I got first interested in the field of scattering amplitudes. Then I have to thank Eduardo Conde, for having eventually pulled me into this field, and for the enriching and strong relationship of professional collaboration and friendship, on which large part of this work rests on. 
Furthermore, I am grateful to all the participants of the \href{http://www.ulb.ac.be/sciences/ptm/pmif/Rencontres/ModaveXII/lectures.html}{XII Modave Summer School in Mathematical Physics}, for being a patient as well as attentive audience. I finally thank C\'eline Zwikel, Eduardo Conde, Riccardo Argurio, Roberto Oliveri, and Victor Lekeu for reading the final drafts of these notes, and returning precious feedback.\\ 
This work is supported by IISN-Belgium (convention 4.4503.15).
\bigskip
\bigskip

\pagestyle{fancy}
\fancyhead{}
\rhead[\fancyplain{}{\scshape\leftmark}]{\fancyplain{}{\thepage}}
\lhead[\fancyplain{}{\thepage}]{\fancyplain{}{\scshape\rightmark}}
\bibliography{OSamplisrefs}
\bibliographystyle{utphys}

%\begin{thebibliography}{99}
%	\bibitem{...} ....
%\end{thebibliography}
	
\end{document}